\documentclass{article}
\usepackage{fleqn,ams}
\usepackage{amssymb}
\usepackage{epsfig}
\usepackage{verbatim}

\newcommand{\REFigure}[2]{%
\begin{center}\epsfig{file=#1,width=7cm,height=7cm,angle=-90}\\[12pt]
\refstepcounter{figure}Figure \thefigure: {\sl #2}\end{center}}
\newcommand{\EFigure}[2]{%
\begin{center}\epsfig{file=#1,width=7cm,height=7cm}\\[12pt]
\refstepcounter{figure}Figure \thefigure: {\sl #2}\end{center}}
\begin{document}

\begin{center}
{\bf\Large  A Kinetics Of The Non-Equilibrium Universe. II. A
Kinetics Of The Local Thermodynamical Equilibrium's Recovery.}\\
Yu.G.Ignatyev, D.Yu.Ignatyev\\
Kazan State Pedagogical University,\\ Mezhlauk str., 1, Kazan
420021, Russia
\end{center}

\begin{abstract}
It has been researched the kinetics of the thermal equilibrium's
establishment in an early Universe under the assumption of the
recovery of interaction scaling of elementary particles in range
of superhigh energies. The case of the thermal equilibrium's weak
initial violation and basic cosmological consequences of the
thermal equilibrium's violation have been researched.
\end{abstract}

\section{Introduction}
In the previous paper of one of the authors \cite{LTE} it was
shown, that in the case of the scaling behavior of the particles'
cross-section of interaction in range of superhigh energies:
\begin{equation}\label{1}
\sigma_{tot}\sim \frac{\mbox{Const}}{s},
\end{equation}
where $s$ is a kinematic invariant of the four-particle reaction
(details see in \cite{LTE}), the initial particles' distribution
in the expanding Universe is not to be equilibrium, but can be
random. In this paper we investigate the kinetics of the processes
with elementary particles in the early Universe under the
conditions of scaling of interactions with the purpose to clarify
the boundaries of randomness of the initial particles'
distribution. As the cross-section of elementary particles'
interaction at that we will use an asymptotic cross-section of
scattering, UACS, incorporated in papers \cite{UACS},
\cite{Yuneq}\footnote{As in the previous paper we will use a
system of units $G=\hbar=c=1$.}:
\begin{equation}\label{2}
\sigma_0(s)=\frac{2\pi}{s\left(1+\ln^2\frac{s}{s_0}\right)}=\frac{2\pi}{s\Lambda(s)},
\end{equation}
where $s_0=4$ - the square of the total energy of two colliding
Planck masses,
\begin{equation}\label{3}
\Lambda(s)=1+\ln^2\frac{s}{s_0}\approx \mbox{Const}.
\end{equation}

\section{Kinetic equations for superthermal particles}
\subsection{A simplification of the relativistic integral of collisions}
The relativistic kinetic equations for homogenous isotropic
distributions $f_a(t,p)$ have the form (see the previous paper
\cite{LTE}, details see in \cite{Yukin3}, \cite{Yukin3_1}):
\begin{equation}\label{4}
\frac{\partial f_a}{\partial t}-\frac{\dot{a}}{a} p\frac{\partial
f_a}{\partial p}=\frac{1}{\sqrt{m_a^2+p^2}}\sum\limits_{b,c,d}^{}
J_{ab\leftrightarrows cd}(t,p),\end{equation}
where $a(t)$ is a scale factor of the Friedmann's world:
\begin{equation}\label{5}
ds^2=dt^2-a^2(t)[d\chi^2+\rho^2(\chi)(\sin^2\theta
d\varphi^2+d\theta^2)];
\end{equation}
\begin{equation}\label{6}
p^2=-g_{\alpha\beta}p^\alpha p^\beta ,
\quad(\alpha,\beta=\overline{1..3});
\end{equation}
$J_{ab\leftrightarrows cd}(t,p)$ is an integral of four-particle
reactions \cite{Yukin1}, \cite{Yukin2}:
$$
J_{ab}(t,p)=\pi^4 \int d\pi_b d\pi_c d\pi_d \delta^{(4)}
(P_a+P_b-P_c-P_d)\times $$
\begin{equation}\label{7}\times[(1\pm f_a)(1\pm f_b)|f_cf_d \overline{M_{cd\to ab}|^2}-
\end{equation}
$$-
(1\pm f_c)(1\pm f_d)f_af_b|\overline{M_{ab\to cd}|^2}],
$$
characters $\pm$ correspond to bosons ($+$) and fermions ($-$),
$M_{i\to f}$ are invariant amplitudes of scattering (line means
the average by particles' states of polarization), $d\pi_a$ is a
normed differential of volume of the $a$ particle's momentum
space:
\begin{equation}\label{8}
d\pi_a=\sqrt{-g}\frac{\rho_a dp^1dp^2dp^3}{(2\pi)^3p_4},
\end{equation}
$\rho_a$ is a factor of degeneration.

Let us simplify an integral of four-particle interactions
(\ref{7}), using the properties of the distributions' isotropy
$f_a(t,p)$. For the fulfilment of two inner integrations we will
proceed to the local c.m. system, integration in which is carried
out by the elementary way. After the inverse Lorentz
transformation and conversion to the spherical system of
coordinates in the momentum space we obtain (\cite{Yudiffuz}):
$$J_{ab}(p)=-\frac{2S_b+1}{8(2\pi)^4p}\int\limits_0^\infty
\frac{qdq}{\sqrt{m^2_b+q^2}}\times $$
$$\times
\int\limits_{s_-}^{s_+} \frac{sds}{s+m_a^2-m_b^2}
\frac{1}{16\pi\lambda^2}\int\limits_{-\lambda^2/s}^0 dt
\overline{|M(s,t)|^2}\int\limits_0^{2\pi}d\varphi\times
$$
$$\{f_a(p_4)f_b(q_4)[1\pm f_c(p_4-\Delta)][1\pm f_d(q_4+\Delta)]-
$$
$$-f_c(p_4-\Delta)f_d(q_4+\Delta)[1\pm f_a(p_4)][1\pm f_b(q_4)]\},$$
where
$$\Delta=-\frac{ts}{\lambda^2}\left[p_4-q_4-(p_4+q_4)\frac{m^2_a-m^2_b}{s}\right]-$$
$$-
\cos\varphi\sqrt{-\frac{ts}{\lambda^2}\left(1+\frac{ts}{\lambda^2}\right)}\times
$$
$$\times
\left[4p_4q_4\left(1-\frac{m^2_a+m^2_b}{s}\right)-\frac{\lambda^2+4m^2_bp^2_4+4m^2_aq^2_4}{s}
\right]^{1/2},$$
$$
\lambda^2=s^2-2s(m^2_a+m^2_b)+(m^2_a-m^2_b)^2,-
$$
a function of a triangle, $s,t$ are the kinematic invariants (see
\cite{LTE}),
$$s_\pm=m^2_a+m^2_b+2(p^4q^4\pm pq).$$
It is necessary at that to bear in mind the definition of the
total cross-section of an interaction \cite{Pilk}, (see also the
previous paper \cite{LTE}):
\begin{equation}\label{9}
\sigma_{tot}=\frac{1}{16\pi
\lambda^2}\int\limits_{-\lambda^2/s}^0dt F(s,t),
\end{equation}
where we denoted as is generally accepted:
\begin{equation}\label{10}
F(t,s)=\overline{|M(s,t)|^2}.
\end{equation}

In the ultrarelativistic limit:
\begin{equation}\label{11}
\frac{p_i}{m_i} \to\infty \Rightarrow \frac{s}{m^2_i}\to\infty;
\quad \lambda\to s^2,
\end{equation}
aforecited expressions are greatly simplified:
$$J_{ab}(p)=-\frac{(2S_b+1}{32(2\pi)^4p}\int\limits_0^\infty
dq \int\limits_{0}^{4pq} \frac{ds}{s} \int\limits_{0}^1 dx
F(x,s)\int\limits_0^{2\pi}d\varphi\times
$$
$$
\{f_a(p)f_b(q)[1\pm f_c(p-\Delta)][1\pm f_d(q+\Delta)]-$$
\begin{equation}\label{12}
-f_c(p-\Delta)f_d(q+\Delta)[1\pm f_a(p)][1\pm f_b(q)]\},
\end{equation}
where a new variable was incorporated (see \cite{LTE}):
\begin{equation}\label{13}
x=-\frac{t}{s}
\end{equation}
and
\begin{equation}\label{14}
\Delta=x(p-q)-\cos\varphi \sqrt{x(1-x)(4pq-s)}.
\end{equation}
\subsection{An integral of collisions for a distributions' weak deviation from the equilibrium}%
Let us investigate at first a weak violation of the
thermodynamical equilibrium in the hot model, when the main part
of particles, $n_0(t)$, lies in the thermal equilibrium state, and
only for the minor part of particles, $n_1(t)$ the thermal
equilibrium is violated (see Fig. \ref{Fig_df}):
\begin{equation}\label{15}
n_1(t)\ll n_0(t).
\end{equation}
\EFigure{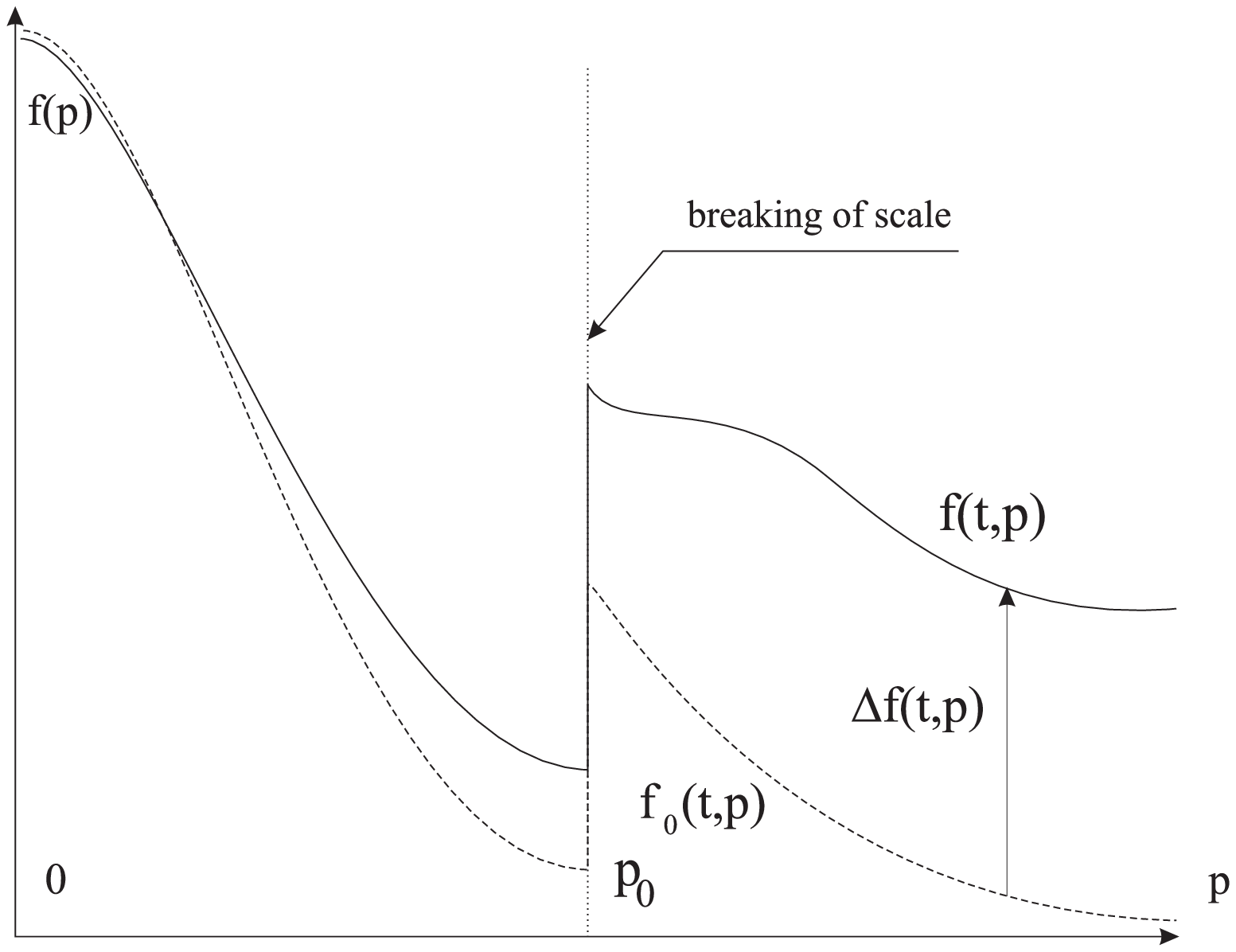}{The schematic representation of the distribution
function's deviation from the equilibrium.\label{Fig_df}}
Henceforth in this paper we will suppose that distribution
functions different not greatly from the equilibrium ones in the
range of small values of energy, smaller than certain unitary
limit, $p=p_0$ (or $T=T_0$), below which scaling is absent, and
can be violated at energies, above the limit:
\begin{equation}\label{16}
\!\!\!\!f_a(p)\approx\left\{\begin{array}{ll}%
\!\!\!f^0_a={\displaystyle
\frac{1}{{\displaystyle\exp(\frac{-\mu_a +E_a(p)}{T})
\pm 1}}}, & p< p_0;\\
 & \\
\!\!\! \Delta f_a(p); \;f_a^0(p)\ll \Delta f_a(p)\ll 1, & p>p_0,\\
\end{array}\right.
\end{equation}
where $\mu_a(t)$ are the chemical potentials, $T(t)$ is a
temperature of the equilibrium component of plasma. Thus, in range
$p>p_0$ it can be observed the anomaly great number of particles
in comparison with the equilibrium one, but minor at that (ñì.
(\ref{15})) in comparison with the total number of equilibrium
particles.

Let us investigate the process of relaxation of the distribution
$f_a(p)$ to the equilibrium $f^0_a(p)$. The problem in such a
setting for the special case of the initial distribution
$f(t=0,p)$ was solved earlier in \cite{UACS}, \cite{Yuneq}. Here
we will give the general solution of this problem. At that, as it
will be obvious from the further arguments, the cosmological
plasma can be formally considered as a two-component system-
equilibrium with the distribution $f_a^0(t,p)$, and nonequilibrium
{\it superthermal}, with the distribution $\delta
f_a(t,p)=\Psi(t,p)$, where number of particles in the
nonequilibrium component is small, but energy density, as a matter
of fact, is random.

Let us investigate an integral of collisions (\ref{12}) in range
\begin{equation}\label{17}p\geq p_0\gg T.
\end{equation}
In consequence of the inequality (\ref{16}) in this range we can
neglect collisions of superthermal particles between themselves,
not going beyond the account of the superthermal particles'
scattering on equilibrium particles. Therefore the value of one of
momentums in the integral of collisions, $P'=p-\Delta$, or
$q'=q+\Delta$ must lie in a thermal range, another's value - in a
superthermal one, outside the unitary limit. A subintegral value
of the integral of collisions is extremely small outside this
range. In consequence of this circumstance we can neglect the
second member in curly brackets (\ref{12}), since it can compete
with the first in asymptotically small variation ranges of
variables $x$ and $\varphi$: $x(1-x)\lesssim T/p\to 0$. The
statistical factors of type $[1\pm f_a(p')]$ in the first member
of an integral (\ref{12}) can noticeably differ from one besides
only in the range of momentums' thermal values. As a result the
integral of collisions (\ref{12}) in the investigated range of
momentums' values can be written down in a form:
$$
\left.J_{ab\leftrightarrow cd}(p)\right|_{p\geq
p_0}=\frac{(2S_b+1)\Delta f_a(p)}{(2\pi)^3p}\times
$$
\begin{equation} \label{18}
\times\int\limits_0^\infty \frac{qf_b^0(q)dq}{\sqrt{m^2_b+q^2}}
\int\limits_{2p(q^4-q)}^{2p(q^4+q)}\frac{ds}{16\pi}\int\limits_0^1
dxF(x,s).
\end{equation}
Using the definition of the total cross-section of scattering
(\ref{12}), we obtain from (\ref{18}):
$$\left.J_{ab\leftrightarrow cd}(p)\right|_{p\geq
p_0}=$$
\begin{equation}\label{19}
=\frac{(2S_b+1)\Delta f_a(p)}{(2\pi)^3p}\int\limits_0^\infty
\frac{qf_b^0(q)dq}{\sqrt{m^2_b+q^2}}
\int\limits_{2p(q^4-q)}^{2p(q^4+q)}\!\!\!\!\!\sigma_{tot}s(s)ds.
\end{equation}
Substituting eventually an expression for $\sigma_{tot}$ in the
inner integral in form of UACS, (\ref{2}), carrying out an
integration with a logarithmic accuracy and summing up the
obtained expression by all channels of reactions, we find finally:
$$\left.J_a(p)\right|_{p\geq
p_0}=$$
\begin{equation}\label{20}
=-\Delta
f_a(p)\sum\limits_b\frac{(2S_b+1)\nu_{ab}}{\pi}\int\limits_0^\infty
\frac{q^2f^0_b(q)}{\sqrt{m^2_b+q^2}}\frac{dq}{\Lambda(\bar{s})},
\end{equation}
where $$\bar{s}=\frac{1}{2}pq^4,$$ $\nu_{ab}$ is a number of
channels of reactions, in which $a$ sort particles can participate
$a$.

Let us calculate values of the integral (\ref{20}) in extreme cases. \\
\noindent {\bf A scattering on nonrelativistic particles}. If $b$
sort equilibrium particles are nonrelativistic, i.e., $q\ll m_b$,
the integral (\ref{20}) is reduced to the expression:
$$\left.J_a(p)\right|_{p\geq
p_0}=$$
\begin{equation}\label{21}
=-8\pi^2\Delta
f_a(p)\sum\limits_b\frac{n^0_b(t)}{m_b}\frac{\nu_{ab}}{{\displaystyle
1+\ln^2\frac{pm_b}{2}}}, \quad (m_b>T).
\end{equation}
\noindent {\bf A scattering on ultrarelativistic particles.} If
$b$ sort equilibrium particles are ultrarelativistic, i.e.,
$m_b\ll T$, moreover their chemical potential is small, -
$\mu_b\ll T$, then calculating the integral (\ref{20}) with
respect to the equilibrium distribution (\ref{16}), we find:
$$\left.J_a(p)\right|_{p\geq p_0}=$$
\begin{equation}\label{22}
=-\frac{\pi}{3}\frac{\tilde{ N}T^2(t)}{1+\ln^2Tp/2}\Delta
f_a(p),\quad (m_b\ll T, \; \mu_b\ll T),
\end{equation}
where
$$\tilde{{\cal N}}=\frac{1}{2}\left[\sum\limits_B (2S+1)+
\frac{1}{2}\sum\limits_F (2S+1)\right]=N_B+\frac{1}{2}N_F;$$ $N_B$
is a number of sorts of equilibrium bosons, $F$ - fermions.

To estimate contributions of nonrelativstic and relativistic
particles to the integral of collisions in an equilibrium
component, we first will calculate their concentrations. A
concentration of ultrarelativistic particles in the hot model is
resulted from the expression (\ref{16}) for the distribution
function of an equilibrium component by means of the substitution
\begin{equation}\label{23}E(p)=p;\quad \mu_a=0\end{equation}
into the formula for the determination of particles' number
density (see \cite{LTE}):
\begin{equation}\label{24}n_a(t)=
\frac{2S_a+1}{2\pi^2}\int\limits_0^\infty f_a(t,p)p^2dp.
\end{equation}
Thus we find (see \cite{LTE}):
\begin{equation}\label{25}
n_a(t)=\frac{(2S_a+1) T^3}{\pi^2}g_n\zeta(3)
\end{equation}
A concentration of non-relativistic equilibrium particles {\it at
under the assumption of conservation of their number} is violated
in proportion to $a^{-3}(t)$. Therefore in conditions of
equilibrium's weak violation the ratio of nonrelativistic
particles' density to density of relict photons is approximately
constant (since $T \sim a(t)^{-1}$:
\begin{equation}\label{26}
\frac{n_0(t)}{n_\gamma(t)}\approx \mbox{Const}=\delta\sim
10^{-10}\div 10^{-9}.
\end{equation}
Calculating the ratio of contributions to the integral of
nonrelativistic and ultrarelativistic particles' collisions, we
obtain:
\begin{equation}\label{27}
J_{non}/J_{ultra} \sim \frac{24\pi n^0_b}{m_b T^2}= \zeta(3)\delta
\frac{64 T(t)}{\pi m_b}\sim 10^{-9} \frac{T}{m_b},
\end{equation}
- ratio of contributions is small at $T\ll 10^9m_b$ and diminishes
with time. Therefore in future we will neglect the contribution of
nonrelativistic particles in the integral of collisions.

\section{A Relaxation Of A Superthermal Component On Equilibrium Particles}
Substituting the received expression (\ref{22}) for the integral
of collisions into the kinetic equations (\ref{4}), we obtain a
kinetic equation, which describes an evolution of an
ultrarelativistic superthermal component in the equilibrium
cosmological plasma:
\begin{equation}\label{28}
p\left(\frac{\partial \Delta f_a}{\partial
t}-\frac{\dot{a}}{a}p\frac{\partial \Delta f_a}{\partial
p}\right)=-\frac{\pi \tilde{N}}{3}\frac{T^2(t)}{1+\ln^2
pT/2}\Delta f_a.
\end{equation}
Taking into account the fact, that the variable:
\begin{equation}\label{29}
\mathcal{P}=a(t)p,
\end{equation}
is an integral of motion \cite{Yukin3}, we proceed to variables
$t,\mathcal{P}$ in the equation (\ref{28}); for any function
$\Psi(t,p)$ at that the following relation takes place:
\begin{equation}\label{30}
\frac{\partial \Psi(t,p)}{\partial
t}-\frac{\dot{a}}{a}p\frac{\partial \Psi(t,p)}{\partial
p}=\frac{\partial \Psi(t,\mathcal{P})}{\partial t}.
\end{equation}
At such a substitution the kinetic equation (\ref{28}) can be
easily integrated in quadratures. For convenience in future we
will define more exactly the normalization of the variable
$\mathcal{P}$. At that there appears a necessity to compare
values, which are used in the nonequlibrium model, with
corresponding values of the standard cosmological scenario, since
all observed cosmological parameters are interpreted in SCS terms.
We also should keep in mind two {\it synchronous} models of
Universe: the real - nonequilibrium model $\mathcal{M}$ with
macroscopic parameters $P(t)$ and the ideal - equilibrium model
$\mathcal{M}_0$, which in given point of time $t$ possesses
certain macroscopic parameters $P_0(t)$.

\subsubsection*{An Ultrarelativistic Universe}
Let us consider the Universe with the ultrarelativistic equation
of state\footnote{It should be noted, that pressure and momentum
have the same denotations.} (a characteristic of a barotropic line
is $\rho=1/3$):
\begin{equation}\label{31}
\varepsilon = 3p.
\end{equation}
Then according to Einstein's equations an energy density of the
Universe is changed by law:
\begin{equation}\label{32}
\varepsilon a^4=\mbox{Const}; \qquad \varepsilon= \frac{1}{32\pi
t^2},
\end{equation}
and a scale factor is changed by law:
\begin{equation}\label{33}
a(t)\sim t^{1/2}.
\end{equation}
From the other hand, an energy density of an equilibrium plasma is
determined via its temperature by the relation (see \cite{LTE}):
\begin{equation}\label{34}
\varepsilon_0={\cal N} \frac{\pi^2 T^4}{15}.
\end{equation}
Therefore, if the Universe was filled up {\it only} by equilibrium
plasma, its temperature $T_0(t)$ would change by law (see
\cite{LTE}):
\begin{equation}\label{35}
{\cal
N}^{1/4}T_0(t)=\left(\frac{45}{32\pi^3}\right)^{1/4}t^{-1/2}\quad
(\sim a^{-1}) ,
\end{equation}
- here we take into account a possible weak dependence of an
effective number of equilibrium types of particles from time,
${\cal N}(t)$. So, let us define more exactly the formula
(\ref{29}) by following way:
\begin{equation}\label{36}
p=\mathcal{P}{\cal N}^{1/4}T_0(t).
\end{equation}
Such is the meaning of the momentum variable $\mathcal{P}$
according to this formula: {\it to within the numerical factor of
order of one $\mathcal{P}$ is the relation of particles' energy to
their average energy in the same point of time in the locally
equilibrium ultrarelativistic Universe}.

Thus, solving the kinetic equation (\ref{28}) with the account of
relations (\ref{30}) and (\ref{36}), we find its solution:
\begin{equation}\label{37}
\Delta f_a(t,\mathcal{P})=\Delta f^0_a(\mathcal{P})
\exp\left[-\frac{\xi(t,\mathcal{P})}{\mathcal{P}}\int\limits_0^t
\frac{y^2(t')dt'}{\sqrt{t'}} \right],
\end{equation}
where: $$\Delta f^0_a(\mathcal{P})=\Delta f_a(0,\mathcal{P}),$$
is an initial deviation from the equilibrium and there is
incorporated the dimensionless function:
\begin{equation}\label{38}
y(t)=\frac{T(t)}{T_0(t)}
\end{equation}
and the parameter, weakly dependent from variables
$t,\mathcal{P}$:
\begin{equation}\label{39}
\xi(t,\mathcal{P})=\frac{\pi \tilde{\cal{
N}}}{3\sqrt{\mathcal{N}}}\left(\frac{45}{32\pi^3}\right)^{1/4}%
\frac{1}{\Lambda(\mathcal{P}TT_0/2)};
\end{equation}
\begin{equation}\label{40}
\Lambda(x)=1+\ln^2x.
\end{equation}
Values $\mathcal{P}\gg 1$ correspond to the approximation $p\gg
p_T\approx T(t)$.

Since $T(t)$  is a temperature of the equilibrium component of
plasma, and $T_0(t)$ is a temperature of the completely
equilibrium in a given point of time Universe, the following
condition is always fulfilled:
\begin{equation}\label{41}
y(t)\leq 1.
\end{equation}
To be the correct solution of the kinetic equations, the function
$\Delta f_a(t,\mathcal{P})$ has to satisfy in all times to the
integral condition (\ref{15}). Since according to the solution
(\ref{36}) the distribution function's $\Delta f_a(t,\mathcal{P})$
deviation from the equilibrium strictly diminishes with time, for
the validity of the solution (\ref{36}) it is sufficient to the
function $\Delta f_a(t,\mathcal{P})$ to satisfy the condition
(\ref{15}) in the initial point of time. This gives:
\begin{equation}\label{42}
\int\limits_0^\infty \Delta f^0_a(\mathcal{P})\mathcal{P}^2d
\mathcal{P}\gg \frac{2}{{\cal N}^{3/4}}y^3_0,
\end{equation}
where $y_0=y(0)\leq 1$.

As an example we will consider a relaxation of a superthermal
component at the initial distribution in form of a staircase
function for the density of particles' number:
\begin{equation}\label{43}
\Delta f^0(\mathcal{P})=\left\{\begin{array}{ll}
{\displaystyle
\frac{\pi^2 \Delta\tilde{N}}{\mathcal{P}_0\mathcal{P}^2}},& \mathcal{P}\leq \mathcal{P}_0;\\
  &  \\
 0,& \mathcal{P}\> \mathcal{P}_0;\\
 \end{array}\right.,
\end{equation}
so that:
\begin{equation}\label{44}
\Delta\tilde{N}=\frac{1}{\pi^2}\int\limits_0^\infty \Delta
f^0(\mathcal{P})\mathcal{P}^2 d\mathcal{P}
\end{equation}
is an initial conformal density of the non-equilibrium particles'
number. On Fig. \ref{Ris2} an evolution of the superthermal
component for such a distribution, at that we laid $y(t)\equiv 1$
is shown .

\REFigure{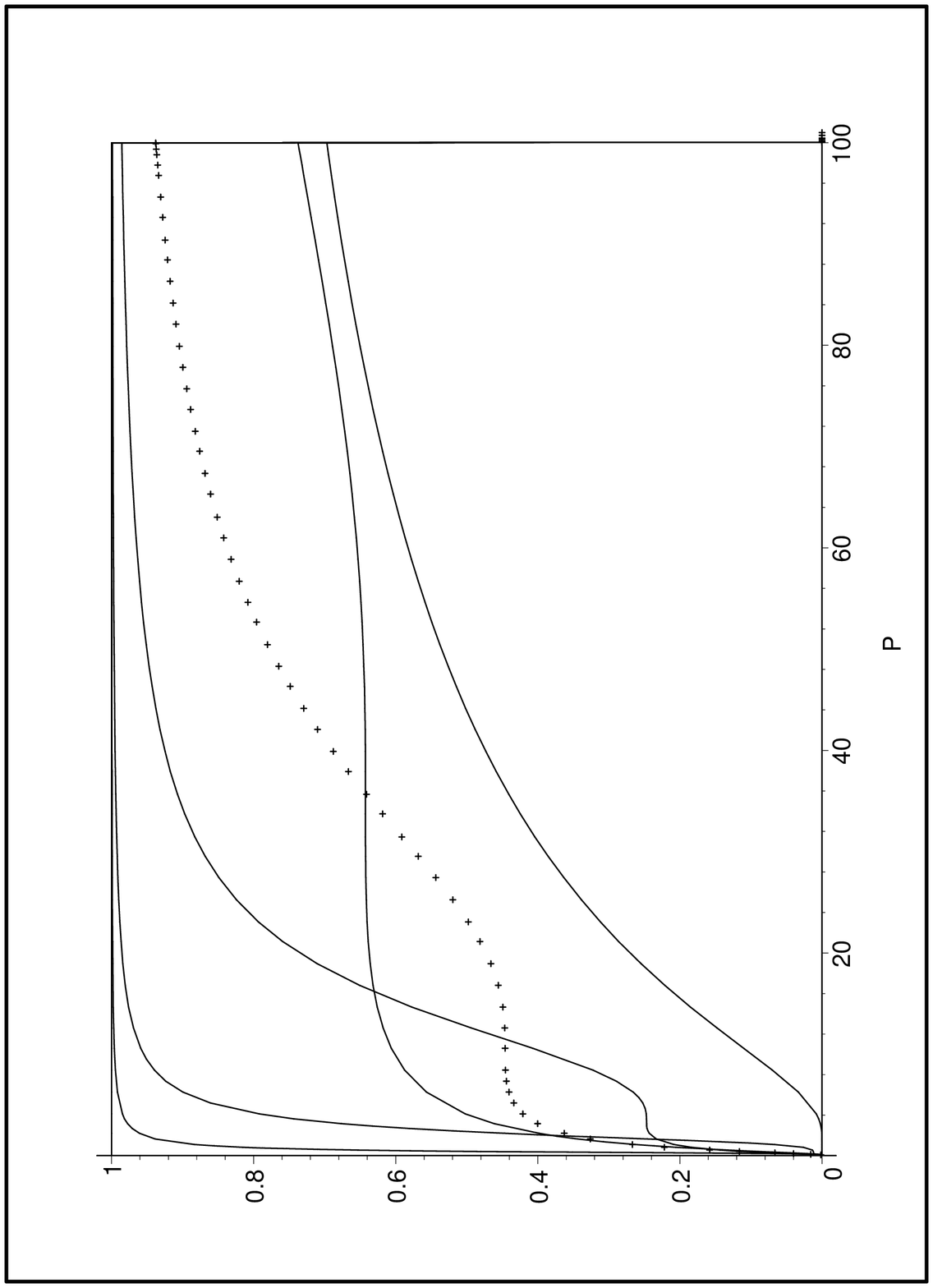}{A relaxation of a superthermal component for
the distribution (\ref{43}) with the assumption $y(t)=1$ at
$\mathcal{P}_0=100$, $\tilde{{\cal N}}/\sqrt{{\cal N}}=10$. A
relative magnitude of the distribution function of particles'
number density by conformal energies $\mathcal{P}$ is shown.
Top-down (along the figure's left border) there are firm lines:
$t=0$, $t=0,01$, $t=0,1$, $t=1$, $t=10$ and $t=100000$; a dotted
line is $t=3$. Time is measured in seconds. \label{Ris2}}

Let us remind that cosmological time $t$ is measured in Planck
units. Therefore a native question, if methods of classical
(nonquantized) kinetics can be used in times of order of several
Planck times, appears. A condition of applicability of a
particles' semi-classical description in the cosmological
situation is a relation, resulting from the Heisenberg's
uncertainty relation:
\begin{equation}\label{45}
Et\gg 1.
\end{equation}
According to (\ref{35}) and (\ref{36}):
\begin{equation}\label{46}
E=p=\mathcal{P}\left(\frac{45}{32\pi^3}\right)^{1/4}t^{-1/2}.
\end{equation}
Therefore a condition of applicability of a particles'
semi-classical description (\ref{45}) takes form:
\begin{equation}\label{47}
t\mathcal{P}^2\gg \sqrt{\frac{32\pi^2}{45}}\approx 2,65.
\end{equation}
Such consideration at $t\sim1$ is justified for sufficiently great
values of the conformal momentum $\mathcal{P}\gg 1$, which exactly
correspond superthermal particles. Thus, {\it a semi-classical
description of particles is applicable in Planck times of an
evolution of the Universe with the more validity the lesser is a
ratio of a thermal energy to a particle's energy}. Thus, a
description of a superthermal (non-equilibrium) component's
evolution does not require a quantum consideration.

\subsubsection*{Non-equilibrium Universe}
Let us consider now the Universe on non-relativistic stage of
expansion. In this case an equation of state is
\begin{equation}\label{48}
p=0
\end{equation}
and a scale factor is changed by law:
\begin{equation}\label{49}
a(t)\sim t^{2/3}.
\end{equation}
We will consider at that a scattering of superthermal particles on
equilibrium massless relic particles, a number of which in SCS is
greater than a number of non-relic particles approximately in
$10^9$. Let $t_0$ is a moment of an ultrarelativistic equation of
state's replacement by a non-relativistic one in the
non-equilibrium Universe and $T^0_\gamma(t)$ is a temperature of
relic photons in the equilibrium Universe. Then:
\begin{equation}\label{50}
T^0_\gamma(t)=\left(\frac{45}{32\pi^3\mathcal{N}}\right)^{1/4}\frac{t^{1/6}_0}{t^{2/3}}.
\end{equation}
Integrating the kinetic equation (\ref{28}), we find in this case:
\begin{equation}\label{51}
\!\!\!\!\!\!\Delta f_a(t,\mathcal{P})\!\!=\!\Delta
f_a(t_0,\mathcal{P})\exp\left[-\frac{\chi(t,\mathcal{P})}{\mathcal{P}}\int\limits_{t_0}^t
\frac{y^2(t')dt'}{t'^{2/3}} \right],
\end{equation}
where $\Delta f_a(t_0,\mathcal{P})$ is determined by the solution
(\ref{37}), and
\begin{equation}\label{52}
\chi(t,\mathcal{P})={\displaystyle \frac{\pi
\tilde{\mathcal{N}}T^0_\gamma(t_0)t^{2/3}_0}{\Lambda(\frac{1}{2}\mathcal{P}T^0_\gamma(t))}}
\end{equation}
is a slowly changing parameter, $\tilde{\mathcal{N}}$ is
determined for massless relic particles.

\section{A Heating Of An Equilibrium Component By Superthermal Particles}
\subsection{An Energy-balance Equation}
In spite of a small in comparison with an equilibrium number of
non-equilibrium particles, energy, included in the non-equilibrium
tail, can turn to be sufficiently greater than an energy of an
equilibrium component, if sufficient distortions of the
distribution lie in a superthermal range, to which great values of
the momentum variable $\mathcal{P}\gg 1$ correspond. Superthermal
particles, colliding with equilibrium ones, transfer theirs energy
to equilibrium particles and thereby heat up an equilibrium
component of plasma. For the finding of the real temperature,
$T(t)$, of an ultrarelativistic plasma with the account of its
heating by superthermal particles we will use the equation
(\ref{32}), which presents itself a consequence of the energy's
conservation law and determines the dependence of the
ultrarelativistic Universe's energy density, $\varepsilon$, from
the cosmological time. This energy density is composed from the
energy density of an equilibrium plasma, $\varepsilon_0$
(\ref{33}), and the energy density of a superthermal component,
$\varepsilon_1$:
\begin{equation}\label{53}
\varepsilon=\varepsilon_0+\varepsilon_1,
\end{equation}
where
\begin{equation}\label{54}
\varepsilon_1=\frac{{\cal N}T^4_0}{\pi^2}\sum\limits_a
(2S_a+1)\int\limits_0^\infty \Delta
f_a(t,\mathcal{P})\mathcal{P}^3d \mathcal{P}.
\end{equation}
For the further it is convenient to incorporate a dimensionless
variable $\sigma(t)$:
\begin{equation} \label{55}
\sigma(t)=\frac{\varepsilon_0}{\varepsilon}=\frac{\varepsilon_0}{\varepsilon_0+\varepsilon_1}\leq
1.
\end{equation}
Since the total energy density $\varepsilon(t)$ is determined from
the other hand via fourth degree of temperature, $T_0(t)$, of the
completely equilibrium Universe, and an energy density of an
equilibrium component $\varepsilon_0(t)$ is determined via fourth
degree of temperature $T(t)$, an incorporated dimensionless
variable is concerned by a simple relation with a dimensionless
variable $y(t)=T(t)/T_0(t)$, incorporated earlier (\ref{38}):
\begin{equation}\label{56}
\sigma(t)=y^4(t),
\end{equation}
from which right away the inequality (\ref{41}) follows.

Thus from (\ref{53}) subject to the solution (\ref{37}) for the
non-equilibrium distribution function we obtain an integral
equation relative to function $y(t)$:
$$
y^4+\frac{15}{\pi^4}\sum\limits_a (2S_a+1) \int\limits_0^\infty
\mathcal{P}^3\Delta f^0_a(\mathcal{P})\times
$$
\begin{equation}\label{57}
\times\exp\left[-\frac{\xi(t,\mathcal{P})}{\mathcal{P}}\int\limits_0^t\frac{y^2(t')dt'}{\sqrt{t'}}
\right]d \mathcal{P}=1.
\end{equation}
From this equation in zero point of time we obtain the relation:
\begin{equation}\label{58}
\frac{15}{\pi^4}\sum\limits_a (2S_a+1) \int\limits_0^\infty
\mathcal{P}^3\Delta f^0_a(\mathcal{P}) d \mathcal{P}=1-\sigma_0.
\end{equation}

\subsection{A Solution Of An Energy-balance Equation}
At specified functions $\Delta f^0_a(\mathcal{P})$ an equation
(\ref{57}) is always integrated with a logarithmic accuracy in
quadratures. Actually, instead of variable $t$ and function $y(t)$
we incorporate new dimensionless variable $\tau$:
\begin{equation}\label{59}
\tau=\frac{\overline{\xi}}{\overline{\mathcal{P}}_0}\sqrt{t}
\end{equation}
and function $Z(\tau)$:
\begin{equation}\label{60}
Z(\tau)=2\int\limits_0^\tau y^2(\tau')d\tau',
\end{equation}
where $\overline{\xi}=\xi(\overline{\mathcal{P}}_0)$, à
\begin{equation}\label{61}
\overline{\mathcal{P}}_0=\frac{\sum\limits_a
(2S_a+1)\int\limits_0^\infty d \mathcal{P}\mathcal{P}^3\Delta
f^0_a(\mathcal{P})}
{\sum\limits_a (2S_a+1)\int\limits_0^\infty d
\mathcal{P}\mathcal{P}^2\Delta f^0_a(\mathcal{P})}
\end{equation}
is an average value of the momentum variable $\mathcal{P}$ in the
point of time $t=0$. At that we obtain:
\begin{equation}\label{62}
y=\sqrt{\frac{1}{2}Z'_\tau};\quad Z(0)=0; \quad
Z'_\tau(0)=2\sqrt{\sigma_0},
\end{equation}
where $\sigma_0=\sigma(0)=y^2(0)$. Then after an integral's
calculation with a logarithmic accuracy the equation (\ref{57})
subject to the relation (\ref{58}) is reduced to the form
\cite{Yuneq}:
\begin{equation}\label{63}
Z'_\tau=2\sqrt{1-(1-\sigma_0)\Phi(Z)},
\end{equation}
where function $\Phi(Z)$ is incorporated:
$$\Phi(Z)=
\frac{\overline{\mathcal{P}}(t)}{\overline{\mathcal{P}}(0)}=$$
\begin{equation}\label{64}
=\frac{\sum\limits_a (2S_a+1)\int\limits_0^\infty d
\mathcal{P}\mathcal{P}^3\Delta f^0_a(\mathcal{P})
e^{-Z\overline{\mathcal{P}}_0/\mathcal{P}}}
{\sum\limits_a (2S_a+1)\int\limits_0^\infty d
\mathcal{P}\mathcal{P}^3\Delta f^0_a(\mathcal{P})}.
\end{equation}%
Henceforth it will be convenient to proceed to a new dimensionless
momentum variable:
\begin{equation}\label{65}
\rho=\frac{\mathcal{P}}{\overline{\mathcal{P}}_0},
\end{equation}
such that:
\begin{equation}\label{66}
1=\frac{\sum\limits_a (2S_a+1)\int\limits_0^\infty d
\rho\rho^3\Delta f^0_a(\rho)}
{\sum\limits_a (2S_a+1)\int\limits_0^\infty d \rho\rho^2\Delta
f^0_a(\rho)} \Rightarrow \overline{\rho}_0 \equiv 1.
\end{equation}
Then:
$$\Phi(Z)=
\frac{\overline{\rho}(\tau)}{\overline{\rho}(0)}=$$
\begin{equation}\label{67}
=\frac{\sum\limits_a (2S_a+1)\int\limits_0^\infty d
\rho\rho^3\Delta f^0_a(\rho) e^{-Z/\rho}}
{\sum\limits_a (2S_a+1)\int\limits_0^\infty d \rho\rho^3\Delta
f^0_a(\rho)}.
\end{equation}%

It is obvious, that $\Phi(Z)$ is a monotone decreasing function of
$Z$, since it is always fair
\begin{equation}\label{68}\Phi'_Z<0,\qquad Z\in(0,\infty),
\end{equation}
at that $\Phi(0)=1$ and $\Phi(\infty)=0$. From this it follows:
\begin{equation}\label{69}
0\leq \Phi(Z)\leq 1.
\end{equation}
From the definition (\ref{67}) it also follows, that the following
relation is always fair
\begin{equation}\label{70}\Phi''_{ZZ}>0,\qquad (Z\in(0,\infty)),
\end{equation}
therefore graph of function $\Phi(Z)$ is concave. In consequence
of a strict monotonicity and continuous differentiability of
function $\Phi(Z)$ according to the equation (\ref{66}) function
$Z(\tau)$ with its first derivative are monotone increasing
functions of the variable $x$. Integrating the equation
(\ref{66}), we obtain:
\begin{equation}\label{71}
\frac{1}{2}\int\limits_0^Z {\displaystyle
\frac{dU}{\sqrt{1-(1-\sigma_0)\Phi(U)}}}=\tau.
\end{equation}
Stated monotonicity properties of functions $\Phi(Z)$ and
$Z(\tau)$ guarantee us the equation (\ref{71}) at each specified
value $\tau$ has the unique solution $Z(\tau)$. At a computed
value of function $Z(\tau)$ a value of temperature $T(t)$ (or
$y(t)$) of an equilibrium component can be obtained according to
(\ref{65}) and (\ref{66}) by means of the formula:
\begin{equation}\label{72}
y=[1-(1-\sigma_0)\Phi(Z)]^{1/4}.
\end{equation}
Thus, the problem of heating of an equilibrium component of plasma
is formally solved.

It should be noted, that a condition of approximation's
applicability of the LTE's weak violation (\ref{42}) in terms of
incorporated here values $\overline{\mathcal{P}}_0$ and $\sigma_0$
takes the form:
\begin{equation}\label{73}
{\cal N}^{1/4}\sigma^{3/4}_0\overline{\mathcal{P}}_0\gg
1-\sigma_0.
\end{equation}

\subsection{An analysis of a thermal heating's process of establishment}
Let us proceed now to the analysis of the obtained solution.
\subsubsection*{An asymptotic behavior at small times}
First we will consider an asymptotic behavior of solutions at
small cosmological times:
\begin{equation}\label{74}
t\to 0 \Rightarrow \tau\ll 1.
\end{equation}
Expanding a subintegral expression in the right side of (\ref{67})
in Taylor's series by the degrees of smallness of $Z$, we obtain
subject to the definition (\ref{64}) an asymptotic expansion of
function $\Phi(Z)$:
\begin{equation}\label{75}
\Phi(Z)= 1-Z +O^2(Z).
\end{equation}
Substituting (\ref{75}) into the equation (\ref{63}) and
integrating the obtained equation subject to initial conditions
(\ref{62}), we find an asymptotic, at small times, solution:
\begin{equation}\label{76}
Z=\tau\sqrt{\sigma_0}+ \tau^2(1-\sigma_0),
\end{equation}
from which subject to (\ref{62}) we obtain:
\begin{equation}\label{77}
y(t)= \sqrt{\sqrt{\sigma_0}+\tau(1-\sigma_0)} .
\end{equation}
The case $\sigma_0\ll 1$ corresponds to a small energy density of
an equilibrium component in comparison with an energy density of a
non-equilibrium component of plasma in the point of time $t=0$.
According to (\ref{73}) a consideration of this case is justified
for sufficiently great values of $\overline{\mathcal{P}}_0$:
\begin{equation}\label{78}
\overline{\mathcal{P}}_0 \gg \sigma_0^{-3/4}.
\end{equation}
In this case we obtain from (\ref{77}) an asymptotic law of
temperature's variation in the early Universe:
\begin{equation}\label{79}
T(t)=T_\gamma^0(t)
\left(\frac{\overline{\xi}}{\overline{\mathcal{P}}_0}
\right)^{1/2}t^{1/4} \sim t^{-1/4}
\end{equation}
a temperature of plasma falls more slowly than in SCS, but in each
point of time the real temperature is lower than the corresponding
temperature in SCS:
\begin{equation}\label{80}
y(t)\leq 1 \Rightarrow T(t)\leq T_0(t),
\end{equation}
and is compared with the last in a certain point of time
$\overline{t}$. Let us find this time. Substituting (\ref{79})
into the solution (\ref{37}), we find a law of evolution of the
superthermal particles' distribution in the Universe's early
stage:
\begin{equation}\label{81}
\Delta f_a(t,\mathcal{P})=\Delta
f^0_a(\mathcal{P})\exp{\displaystyle \left(-\frac{\overline{\xi}\
^2t}{\overline{\mathcal{P}}_0\mathcal{P}}\right)}.
\end{equation}
According to (\ref{79}) and (\ref{81}) LTE as a whole is recovered
in times
\begin{equation}\label{82}
t>\overline{t}=\left(\frac{\overline{\mathcal{P}}_0}{\overline{\xi}}\right)^2,
\end{equation}
is exactly a time of an equilibrium's establishment. A plasma's
temperature at that reaches its equilibrium value $T_0(t)$
(constantly decreasing at that with time). However, for particles
with energies higher than the average one
$\overline{\mathcal{P}}_0$, the equilibrium is not reached yet in
this time.
\subsubsection*{An asymptotic behavior at great times in the ultrarelativistic Universe}
Supposing $t>\overline{t}$ and, consequently, $y(t)\approx 1$, we
will obtain an asymptotic form (\ref{37}) on great times:
\begin{equation}\label{83}
\Delta f_a(t,\mathcal{P})=\Delta
f^0_a(\mathcal{P})\exp\left(-2\frac{\xi(\mathcal{P})
\sqrt{t}}{\mathcal{P}}\right);
\end{equation}
thus, thermal heating's establishment time for particles with the
momentum $\mathcal{P}$ is:
\begin{equation}\label{84}
t_{\mathcal{P}}=\left[\frac{\mathcal{P}}{\xi(t,\mathcal{P})}\right]^2
\approx
\left(\frac{\mathcal{P}}{\overline{\mathcal{P}}_0}\right)^2t.
\end{equation}

Let us suppose, that in range of great values of $\mathcal{P}$ the
initial distribution of superthermal particles is extrapolated via
power law:
\begin{equation}\label{85}
\Delta f_a^0(\mathcal{P}) \sim \mathcal{P}^{-\lambda};
\quad(\mathcal{P}\gg 1; \; \lambda>4).
\end{equation}
Then an average energy of superthermal particles
$\overline{E}_1(t)$, for which a thermal equilibrium to moment of
time $t$ is not reached yet, is equal:
\begin{equation}\label{86}
\overline{E}_1(t)=\xi(t,\mathcal{P})T_0(t)\sqrt{t}\approx
\frac{1}{\Lambda(T^2_0(t)\overline{\mathcal{P}}_0)}\sim
\mbox{Const}
\end{equation}
practically does not depend on time. In the previous paper
\cite{LTE} we noted that in the modern stage $1/\Lambda \sim
\alpha^2$, where $\alpha$ is a fine structure constant. Thus, we
obtain from (\ref{86}) the estimation:
\begin{equation}\label{87}
\overline{E}_1(t)\sim 10^{-4}
\end{equation}
is in ordinary units $\overline{E}_1(t)\sim 10^{15}$ Gev - i.e,
particles with an energy of order of a grand unification character
energy or higher stay non-equilibrium. This conclusion is
fundamentally important for cosmological scenarios.

\subsubsection*{An asymptotic behavior at great times in the non-relativistic Universe}
Let us consider now an evolution of superthermal ultrarelativistic
particles in the non-relativistic Universe, the thermal
equilibrium in which on average is already recovered. Supposing in
this case $y=1$ in the expression (\ref{51}) for the distribution
function of superthermal particles, we obtain:
$$\Delta f_a(t,\mathcal{P})= \Delta f_a(t_0,\mathcal{P})\times$$
\begin{equation}\label{88}
\times\exp\left\{ -\frac{3(\pi
\mathcal{N})^{1/4}\chi(\mathcal{P})}{\mathcal{P}}\sqrt{t_0}\left[
\left(\frac{t}{t_0}\right)^{1/3}-1 \right] \right\},
\end{equation}
the distribution of superthermal particles evolves more slowly,
than (\ref{83}). A time of the establishment of the thermal
equilibrium for particles with momentum $\mathcal{P}$ in this case
is equal:
\begin{equation}\label{89}
t_\mathcal{P}\simeq
\frac{1}{\sqrt{t_0}}\left[\frac{\mathcal{P}}{\xi(t,\mathcal{P})}\right]^3,
\end{equation}
and an average energy os superthermal particles falls slowly with
time:
\begin{equation}\label{90}
\overline{E}_1(t)\simeq
\frac{\overline{\mathcal{N}}}{\sqrt{\mathcal{N}}\Lambda(T^2_0(t)
\mathcal{P}/2)}\left(\frac{t_0}{t}\right)^{1/3}\sim\end{equation}
$$
\sim 10^{15}\left(\frac{t_0}{t}\right)^{1/3}\mbox{Gev}.
$$
In the modern age this value is of order $10^{12}\div 10^{13}$Gev
at variation of $t_0$ in limits $10^9\div 10^{12}$sec.

\subsubsection*{Numerical model}
Let us proceed now from estimations to numerical calculations. Let
us present an initial deviation of the distribution function from
the equilibrium in the form:
\begin{equation}\label{91}
\Delta
f^0(x)=\frac{A}{\mathcal{P}_0^3(k^2+x^2)^{3/2}}\chi(1-x),\quad
k\rightarrow 0,
\end{equation}
where $\chi(z)$ is a Heaviside's function (a staircase function),
$x=\mathcal{P}/\mathcal{P}_0$ is a dimensionless momentum
variable, $A$, $\mathcal{P}_0$ and $k$ are certain parameters; a
parameter $k$ is introduced for the security of convergence of all
distribution function's moments in the range of momentum's small
values. The distribution function of superthermal particles'
energy density at that has a form, similar to the spectrum of
so-called {\it flat noise}, when all values of energy are
equiprobable (see Fig. \ref{Ris3} and \cite{Diffuz}).

\REFigure{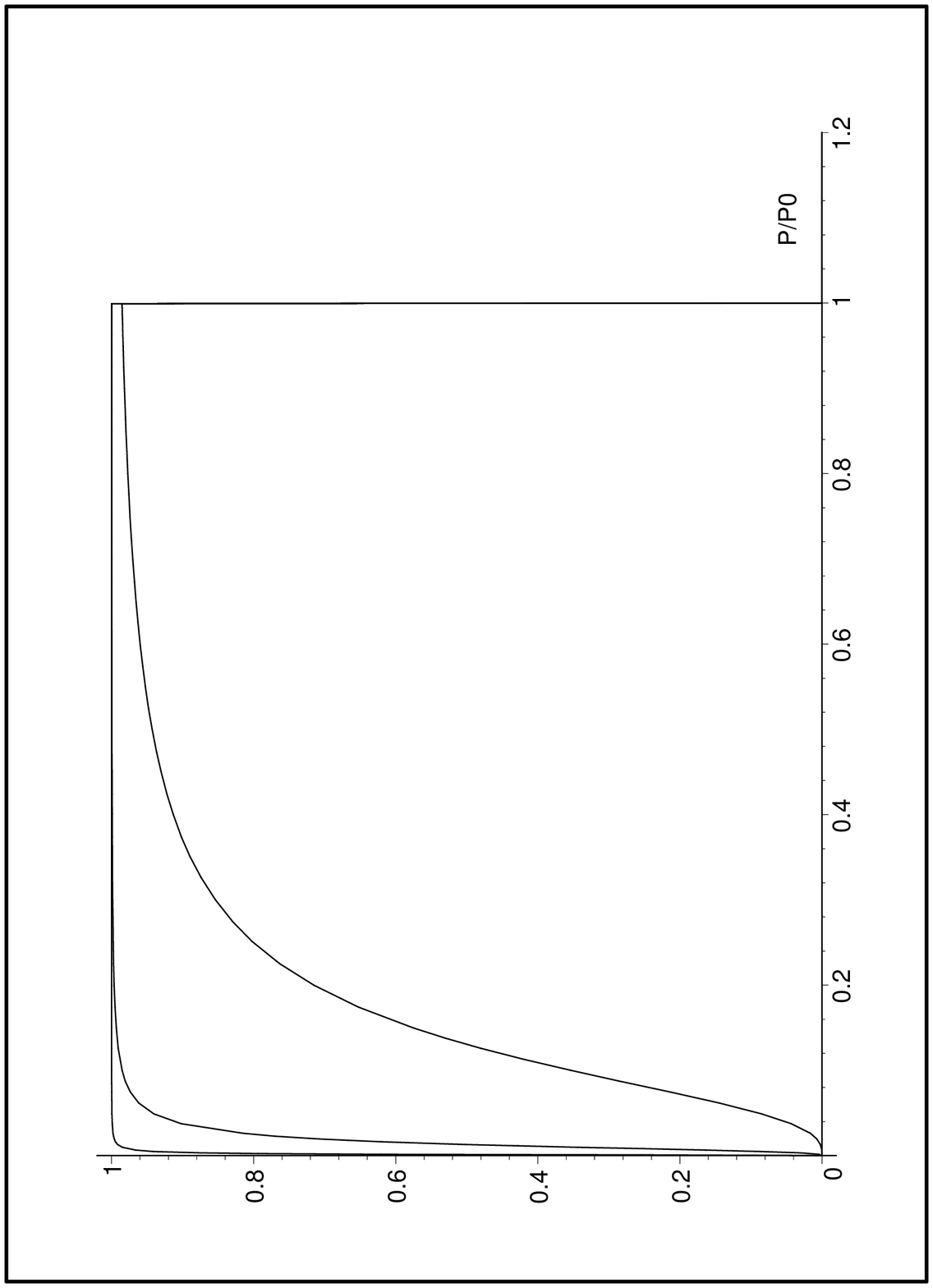}{\label{Ris3} An initial deviation of the
distribution function of the superthermal particles' energy
$\Delta f^0(x)x^3$ from the equilibrium by the formula (\ref{91}),
bottom-up is: $k=0,1$, $k=0,01$, $k=0,001$.}

Calculating a number density of particles relatively to the
distribution (\ref{91}), we find according to (\ref{24}):
$$
n_1(t)=$$ $$=\frac{A(2S_a+1)\mathcal{N}^{3/4}}{2\pi^2}T^3_0(t)
\left(\ln\frac{k}{\sqrt{1+k^2}-1}-\frac{1}{\sqrt{1+k^2}}\right).
$$
Thus, we obtain in the extreme case:
$$
\quad n_1(t) \simeq
A\frac{(2S_a+1)\mathcal{N}^{3/4}}{2\pi^2}T^3_0(t)\ln\frac{2}{k}\qquad
(k\to 0).
$$
from which follows, that for a fulfilment of the condition
(\ref{15}) in all times it is necessary and sufficiently the
following relation to be fulfilled:
$$A\mathcal{N}^{3/4}\ln\frac{2}{k}\ll 2\sigma_0^{3/2}\zeta(3).$$
Calculating now an energy density relatively to the distribution
(\ref{91}), we obtain:
$$\varepsilon_1(t)=
A\frac{\mathcal{N}T^4_0(t)(2S+1)}{\pi^2} \mathcal{P}_0(1-k)^2.$$
Thus, subject to (\ref{36}) we obtain a dependence of a momentum
variable's average value through parameters $\mathcal{P}_0$ and
$k$ (Fig. \ref{Ris4}):
$$\overline{\mathcal{P}}_0=
\mathcal{P}_0\frac{(1-k)^2}{{\displaystyle
\ln\frac{k}{\sqrt{1+k^2}-1}-\frac{1}{\sqrt{1+k^2}}}}$$
\REFigure{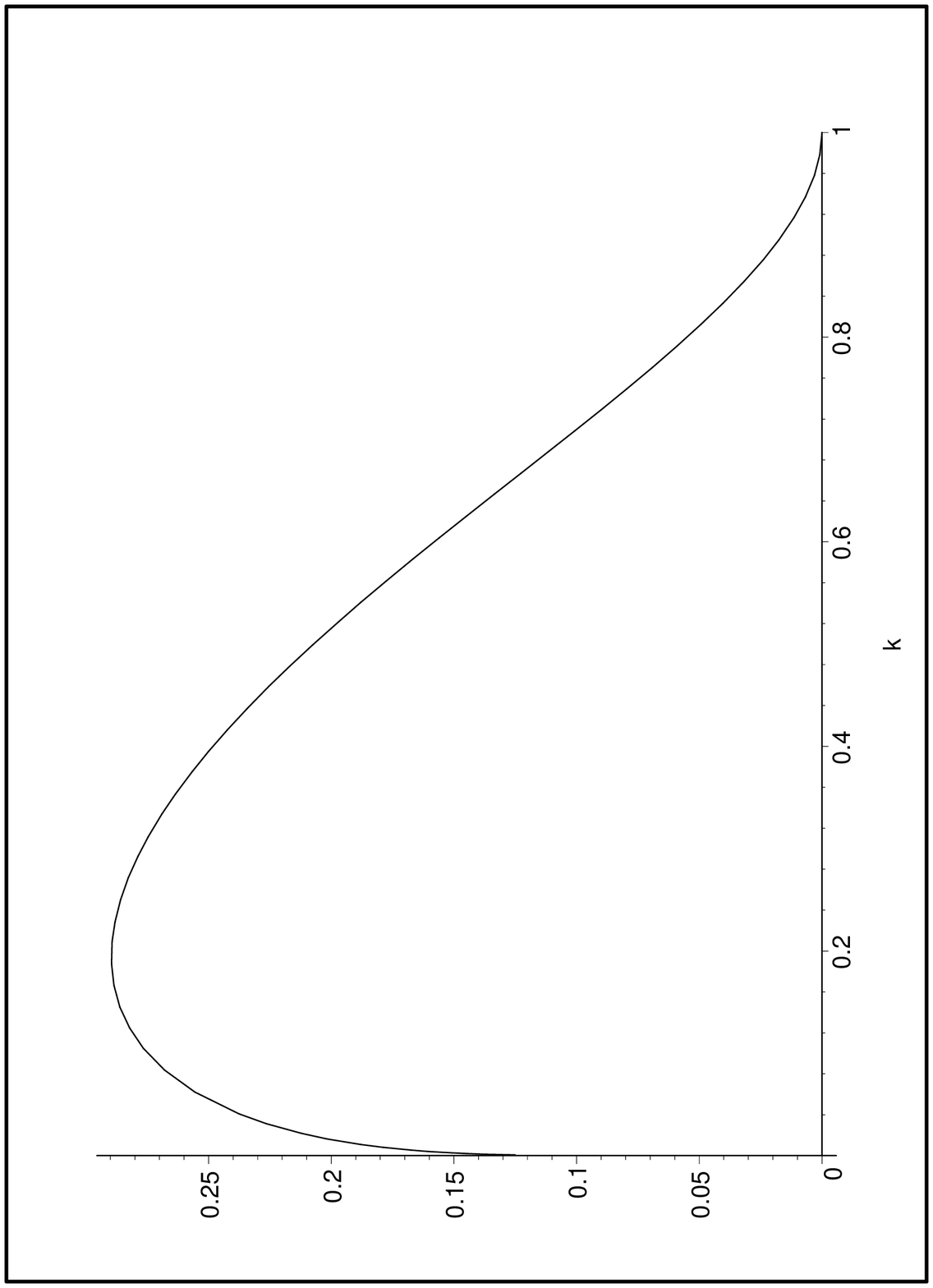}{\label{Ris4}A dependence of variable's relative
average $\overline{\mathcal{P}}_0/\mathcal{P}_0$ from the
parameter $k$ of the distribution (\ref{91}).}

Then an energy-balance equation (\ref{53}) or (\ref{57}) at $t\to
0$ leads to the relation between the distribution parameters
(\ref{91}):
\begin{equation}\label{92}
\frac{15}{\pi^4}A(2S+1)\mathcal{P}_0(1-k)^2 =1-\sigma_0.
\end{equation}
Thus, the following condition, imposing a restriction on model's
parameters, has to be fulfilled:
\begin{equation}\label{93}
A\mathcal{P}_0<\frac{\pi^4}{30},
\end{equation}
Calculating the function $\Phi(Z)$ relatively to the distribution
(\ref{91}) according to the formula (\ref{67}), we find,
proceeding to the limit $k\rightarrow 0$:
\begin{equation}\label{94}
\Phi(Z)=e^{-Z}+\mbox{Ei}(-Z),
\end{equation}
where $\mbox{Ei}(z)$ is an integral exponential function
\cite{Lebed}:
\begin{equation}\label{95}
\mbox{Ei}(z)=\int\limits_{-\infty}^z \frac{e^t}{t}dt, \quad
|\mbox{arg}(-z)|< \pi.
\end{equation}
The equation (\ref{71}) accounting the relation (\ref{95}) takes
the form:
\begin{equation}\label{96}
\tau=\frac{1}{2}\int\limits_0^\theta
\frac{du}{\sqrt{1-(1-\sigma_0)(\exp(-u)+u\mbox{Ei}(-u))}},
\end{equation}
where $\tau=x/\mathcal{P}_0$. If we could integrate (\ref{96}) and
found the dependence $\theta(\tau)\rightarrow Z(x)$, we could
receive the required dependence $y(x)$ from the relations
(\ref{62})-(\ref{63}) in the form:
\begin{equation}\label{97}
y=\left[1-(1-\sigma_0)(e^{-\theta(\tau)}+\theta(\tau)\mbox{Ei}(-\theta(\tau))
\right]^{1/4}.
\end{equation}
However, the dependence $\theta(\tau)$ of course, can not be
found. Therefore we act in the following way: we find the
dependence $\tau(\theta)$ from the equation (\ref{96}) by means of
the direct numerical integration. Then the equation (\ref{97}) can
be considered as an equation of type $y=y(\theta)$, and the set of
two equations (\ref{96}) and (\ref{97}) - as parametric equations
of the graph $y=y(\tau)$:
\begin{equation}\label{98}\!\!\!\!\!
\begin{array}{ll}\\
\tau(\theta) = &\!\!\!\!\!{\displaystyle
\frac{1}{2}\int\limits_0^\theta
\frac{du}{\sqrt{1-(1-\sigma_0)(\exp(-u)+u\mbox{Ei}(-u))}},}\\
 & \\
y(\theta)= & \!\!\!\!\!{\displaystyle
\left[1-(1-\sigma_0)(e^{-\theta(\tau)}+\theta(\tau)\mbox{Ei}(-\theta(\tau))
\right]^{1/4}}
\end{array}
\end{equation}
On Fig. \ref{Ris5} the results of a numerical modelling of the
equilibrium component's heating process by superthermal particles
for the initial distribution (\ref{91}) on the basis of equations
(\ref{98}) are shown.

From this figure it is obvious that at $\tau \sim 1$ temperature
of plasma's equilibrium component in fact becomes saturated up to
the value $T_0(t)$, i.e., to the point of time $t\sim
\mathcal{P}_0^2/\overline{\xi}^2$, that coincides with the
asymptotic estimation (\ref{82}).

\REFigure{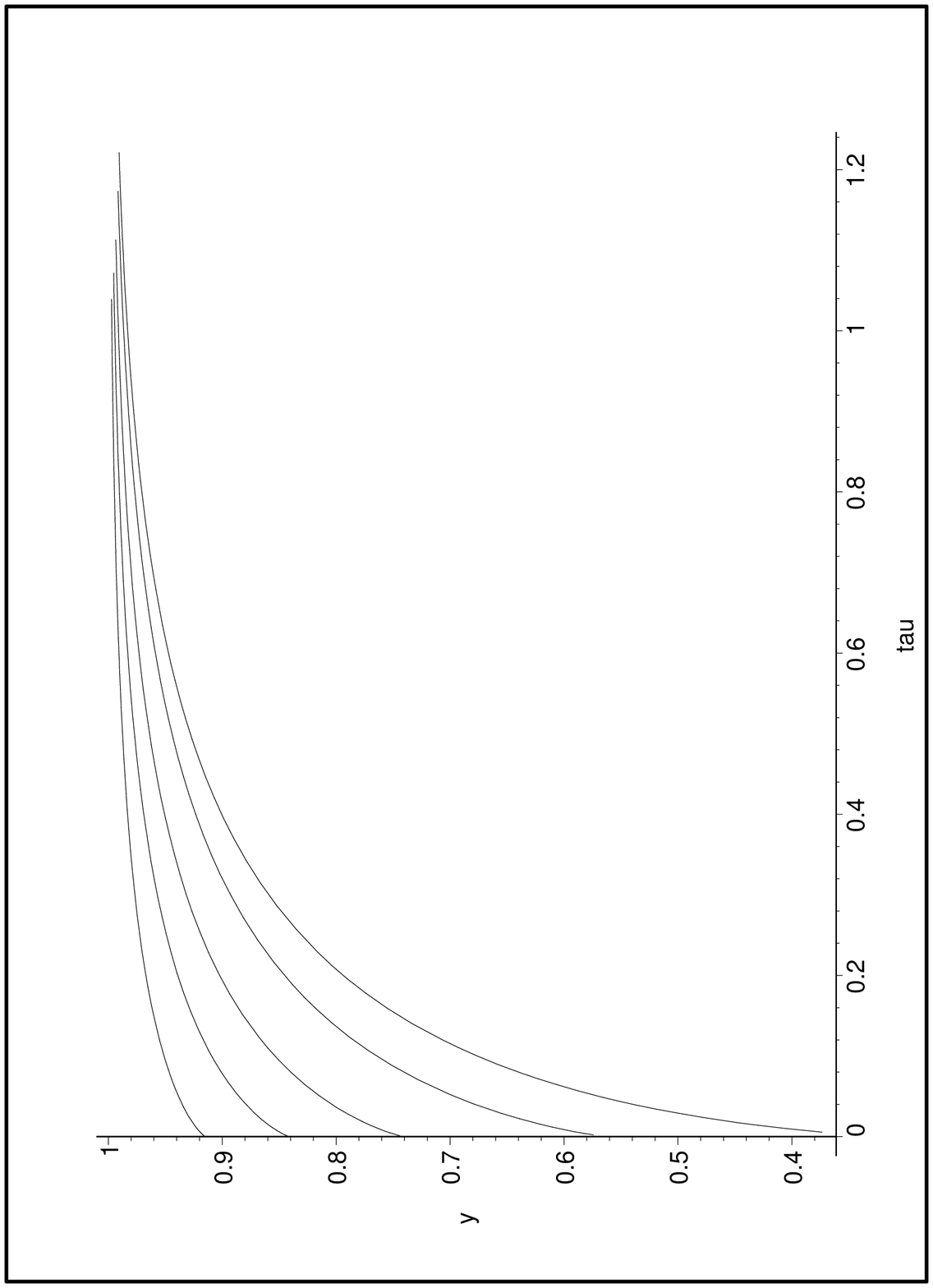}{\label{Ris5}A relaxation of plasma's
temperature to the equilibrium: $y=T(t)/T_0(t)$ subject to the
parameter $\sigma_0$: – bottom-up $\sigma_0= 0,01;\; 0,1;\; 0,3;\;
0,5;\; 0,7$. Values of the dimensionless time variable $\tau$ are
positioned on the abscissa axis.}

\section{Basic regularities of a cosmological model with an initially weakly
violated thermal equilibrium}
\subsection{Concentrations of ultrarelativistic relic particles}
Equilibrium concentrations of ultrarelativistic particles in the
Universe with a weakly violated thermodynamical equilibrium are
determined by the relation (\ref{25}), which subject to the
definition (\ref{38}) can be rewritten in terms of the
incorporated function $y(t)$:
\begin{equation}\label{99}
n_a(t)=\frac{\rho T^3(t)}{\pi^2}g_n\zeta(3)=n^0_a(t)y^3(t)\leq
n^0_a(t),
\end{equation}
where $n^0_a(t)$ are the equilibrium concentrations of the same
particles in SCS:
$$n^0_a(t)=\frac{\rho T_0^3(t)}{\pi^2}g_n\zeta(3)=$$
\begin{equation}\label{100}
=\frac{\rho
}{\pi^2}g_n\zeta(3)\left(\frac{45}{3\pi^3\mathcal{N}}\right)^{3/4}t^{-3/2}.
\end{equation}
Thus, the number of ultrarelativistic particles lying in the
equilibrium in each given point of time $t$ is lesser than in SCS.
If the interaction between particles is described by the scaling
cross-section of type (\ref{1}) or (\ref{2}), than a number of
non-equilibrium particles is described by the distribution
function of type (\ref{37}). Let us consider particles' reactions
in the range of low energies, in which scaling can be violated.
Let $\tau^a_{eff}(t)$ is an effective time of interactions of sort
``$a$'' particles with the other particles and let in the
investigated range of energies the inequality:
\begin{equation}\label{101}
\tau^a_{eff}(t)>t
\end{equation}
has as its solution:
\begin{equation}\label{102}
t>t^*_a.
\end{equation}
Then in the point of time $t=t^*_a$ sort ``$a$'' particles cease
to interact with the other, i.e., become {\it relic}
--- the number of such stable particles in times, later than
$t^*_a$, is conserved, but their density evolves henceforth by
law:
\begin{equation}\label{103}
n_a(t)=n^0_a(t^*_a)y^3(t^*_a)\left[\frac{a(t^*_a)}{a(t)}\right]^3.
\end{equation}
Therefore, if in the point of time $t=t^*_a$ an equilibrium is not
recovered as a whole, the number of relic ``$a$'' sort particles
is in $y^3(t^*_a)$ smaller, than it is received in the standard
scenario\footnote{The time $t^*_a$ is often called the hardening
time (see for example \cite{Zeld}).}. Thus, varying a parameter of
the non-equilibrium Universe's model $\overline{\mathcal{P}}_0$ at
$\sigma_0\ll 0$ we can regulate the number of relic particles and
make this number an arbitrarily small \cite{Yudiffuz} - at a
increase of $\overline{\mathcal{P}}_0$ the number of relic
particles decreases. Thus, at $\overline{\mathcal{P}}_0>10^2$ and
$\sigma_0\ll 1$ relic extra-massive bosons disappear, at
$\overline{\mathcal{P}}_0>3\cdot 10^{17}$ and $\sigma_0\ll 1$
relic neutrinos disappear. This approximate estimator, made in
papers, \cite{Yuneq}, \cite{Yudiffuz}, will be specified below.
\subsection{Relic Neutrinos}
So, let us investigate the problem of relic neutrinos' outcome
during the hardening process. A thermal equilibrium of electron
and muon neutrino is established, generally, by reactions:
\begin{equation}\label{104}
\left.\begin{array}{lll} e^-+ e^+&\rightleftarrows & \nu_e+\overline{\nu}_e,\\
\mu^+ + \mu^- &\rightleftarrows & \nu_\mu+\overline{\nu}_,\\
\mu^+ & \rightarrow & e^++\overline{\nu}_\mu+\nu_e,\\
\mu^- & \rightarrow & e^-+\nu_\mu+\nu_e.\\
\end{array}\right\}
\end{equation}
The cross-section of an electro-weak interaction, corresponding to
reactions of the neutrino annihilation (\ref{104}) in the
interesting for us range of sufficiently low energies $E$, is
described by the expression:
\begin{equation}\label{105}
\sigma_nu\approx \frac{G^2_\mu E^2}{\hbar ^4 c^4},
\end{equation}
where $G_\mu \approx G_nu:=1,4358\cdot 10^{-49}$erg/cm$^3$ is a
constant of an electro-weak interaction. Calculating the hardening
time of electron neutrinos, $t_\nu$ by means of the formula:
\begin{equation}\label{106}
\tau_{eff}=\frac{1}{n_{e}(t_\nu)\sigma_\nu(t_\nu) c}=t_\nu,
\end{equation}
where it is necessary to substitute $E=T(t)$, an expression for an
equilibrium density of ultrarelativistic electrons (\ref{25}) and
a temperature of an equilibrium component $T(t)$, calculated
above. Then the ratio of the electron neutrinos' number after the
hardening in a non-equilibrium model to the same number in an
equilibrium one $N_\nu$, is determined by means of the expression:
\begin{equation}\label{107}
N_\nu=\frac{T(t_\nu)^3}{T_0(t_\nu)^3}=y^{3/4}(t_\nu).
\end{equation}
On Fig. \ref{Ris5} the results of numerical calculations of
electron neutrinos' outcome in a weakly-equilibrium model of the
Universeare are shown. These results, in general, confirm referred
above qualitative assessments of previous papers.

\REFigure{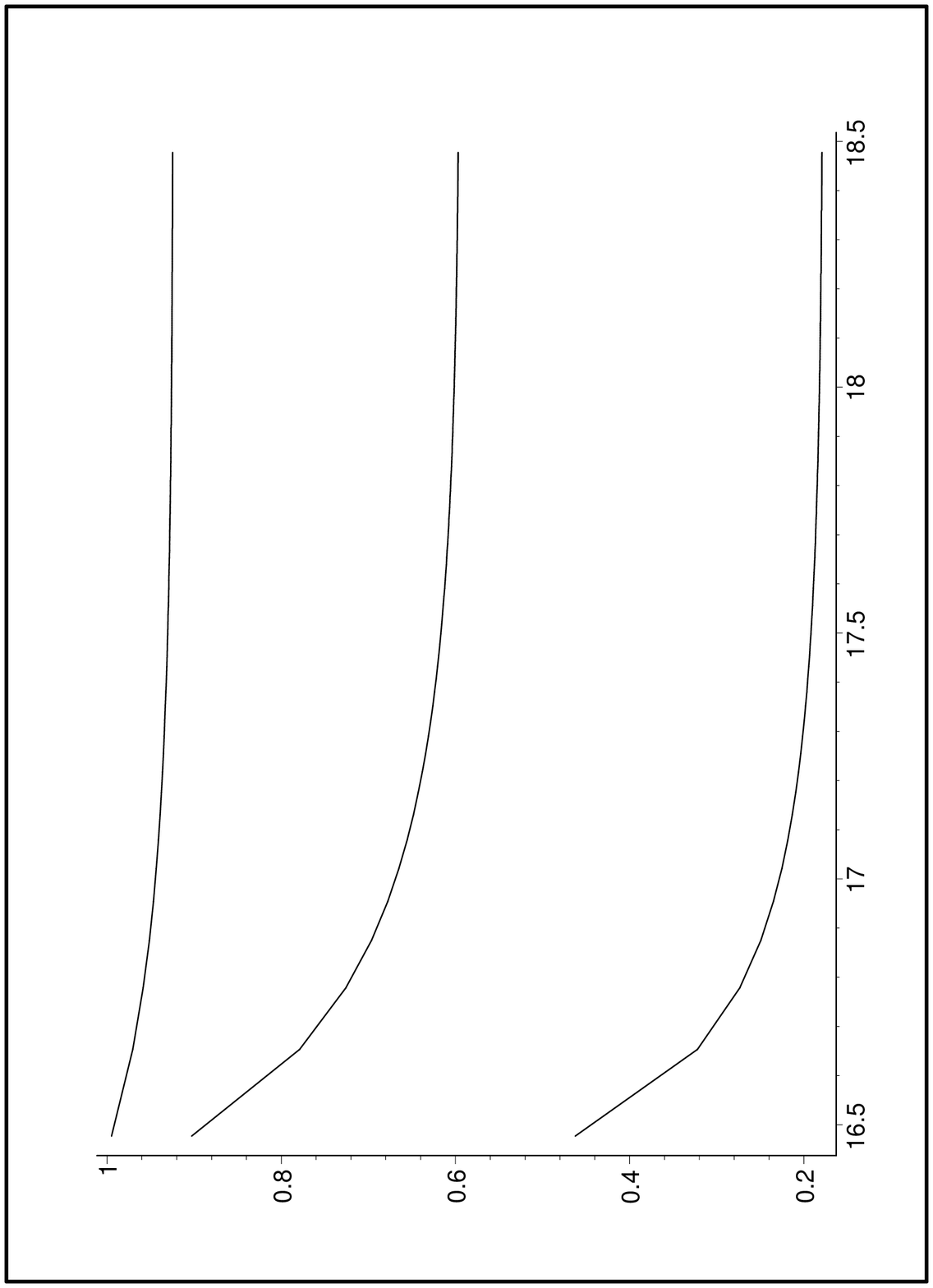}{\label{Ris5}An outcome of electron relic
neutrinos after the hardening relative to a standard model,
$n_\nu/n^0_\nu$, (an ordinate axis) subject to a parameter of a
non-equilibrium model, $lg \mathcal{P}_0$, (an abscissa axis).
Top-down: $\sigma_0=0,9$, $\sigma_0=0,5$, $\sigma_0=0,1$.}

\subsection{A hardening of neutrinos and cosmological helium's generation}
A thermal equilibrium of non-relativistic neutrinos is established
in reactions of type:
\begin{equation}\label{107}
\left.\begin{array}{lll} e^+ +n&\rightleftarrows & p+\overline{\nu},\\
\nu+n &\rightleftarrows &p+e^-.\\
\end{array}\right\}
\end{equation}
A concentration ratio of relic neutrinos, $N_n$, is determined via
the kinetic equation\footnote{Details see, for example, in
\cite{Zeld}.}:
\begin{equation}\label{108}
\frac{dN_n}{dt}+(a+b)N_n=b,
\end{equation}
where:
\begin{equation}\label{109}
N_n(t)+N_p(t)=1,
\end{equation}
($N_p$ is a concentration ratio of protons), coefficients $a(t)$
and $b(t)$ (velocities of reactions (\ref{107})) are related via
the identity:
\begin{equation}\label{110}
b(t)=a(t)e^{-\Delta mc^2/T(t)},
\end{equation}
$\Delta m=m_n-m_p \approx 1,3$Mev is a mass defect of neutron.

Let us investigate the equation (\ref{108}). At $t\to 0$ according
to (\ref{110}) $a(t)\to b(t)$, therefore at $t\to 0$ from
(\ref{105}) we obtain the relation:
\begin{equation}\label{111}
\left.\frac{dN_n}{dt}\right|_{t\to 0}=a(t)[1-2N_n(0)].
\end{equation}
From (\ref{111}) right away follows, that at $N_n(0)>1/2$ the
concentration of neutrons from the very beginning decreases with
time, and at $N_n(0)<1/2$ - increases with time. In the second
case the function $N_n(t)$ always has a maximum in the point of
time $t_*$, determinable by the relation:
\begin{equation}\label{112}
N_n(t_*)={\displaystyle \frac{1}{1+e^{\Delta mc^2/T(t_*)}} }<
\frac{1}{2}.
\end{equation}
At $N_n(0)=1/2$ developing into the Taylor's series $N_n(t)$ by
the small $t$ and using at that the equation (\ref{108}) and
limitary relations at $t\to 0$, we obtain:
$$N_n(t)\approx \frac{1}{2} +
\frac{1}{4}\left.a(t)\frac{\Delta
mc^2}{T^2}\frac{dT}{dt}\right|_{t=0}t^2 +\ldots$$
Since the second member in the right side of the development is
the negative one, then at $N_n(0)=1/2$ the concentration of
neutrons decreases.

Let us proceed now to the numerical modelling. The coefficient
$a(t)$ in units $\sec^{-1}$ at high temperatures is described by
an approximate expression (see for example, \cite{Zeld}):
\begin{equation}\label{113}
a(t)\approx 1,61\cdot \tilde{T}^5(t),
\end{equation}
where $\tilde{T}(t)$ is a temperature in Mev. At low temperatures
$a(t)=W\approx 10^{-3} sec^{-1}$, where $W$ is a probability of
free neutron's decay. Supposing in (\ref{35}) subsequent to
\cite{Zeld} $\mathcal{N}=4,5$, we obtain for $\tilde{T}_0(t)$ the
following relation:
\begin{equation} \label{114}
\tilde{T}_0=0,89 t^{-1/2},
\end{equation}
where time is measured in seconds. How the numerical calculations
show, this point of time is reached sufficiently fast\footnote{The
value of $N_n$ at that in maximum is amount to approximately
0,45.}, - the further history of the nucleosynthesis does not
depend in fact from the initial concentration ratio of neutrons.

Let us consider reactions (\ref{107}) on conditions, when the
equilibrium in plasma is not reached as a whole, i.e., $y(t)<1$.
Supposing at that function $y(t)$ obey the asymptotic form
(\ref{77}), and at interesting for us time scales $\Lambda \simeq
\alpha^{-2} \approx (137)^2$, we obtain:
\begin{equation}\label{115}
\tilde{T}\approx 0,89
\frac{1}{\sqrt{t}}\sqrt{\sqrt{\sigma_0}+0,523(1-\sigma_0)\frac{\sqrt{t}}{P_0}},
\end{equation}
where:
$$P_0=\frac{\mathcal{P}_0}{10^{17}}.$$
On Fig. \ref{Ris6}-\ref{Ris9} the results of the numerical
integration of the equation (\ref{115}) are shown.
\REFigure{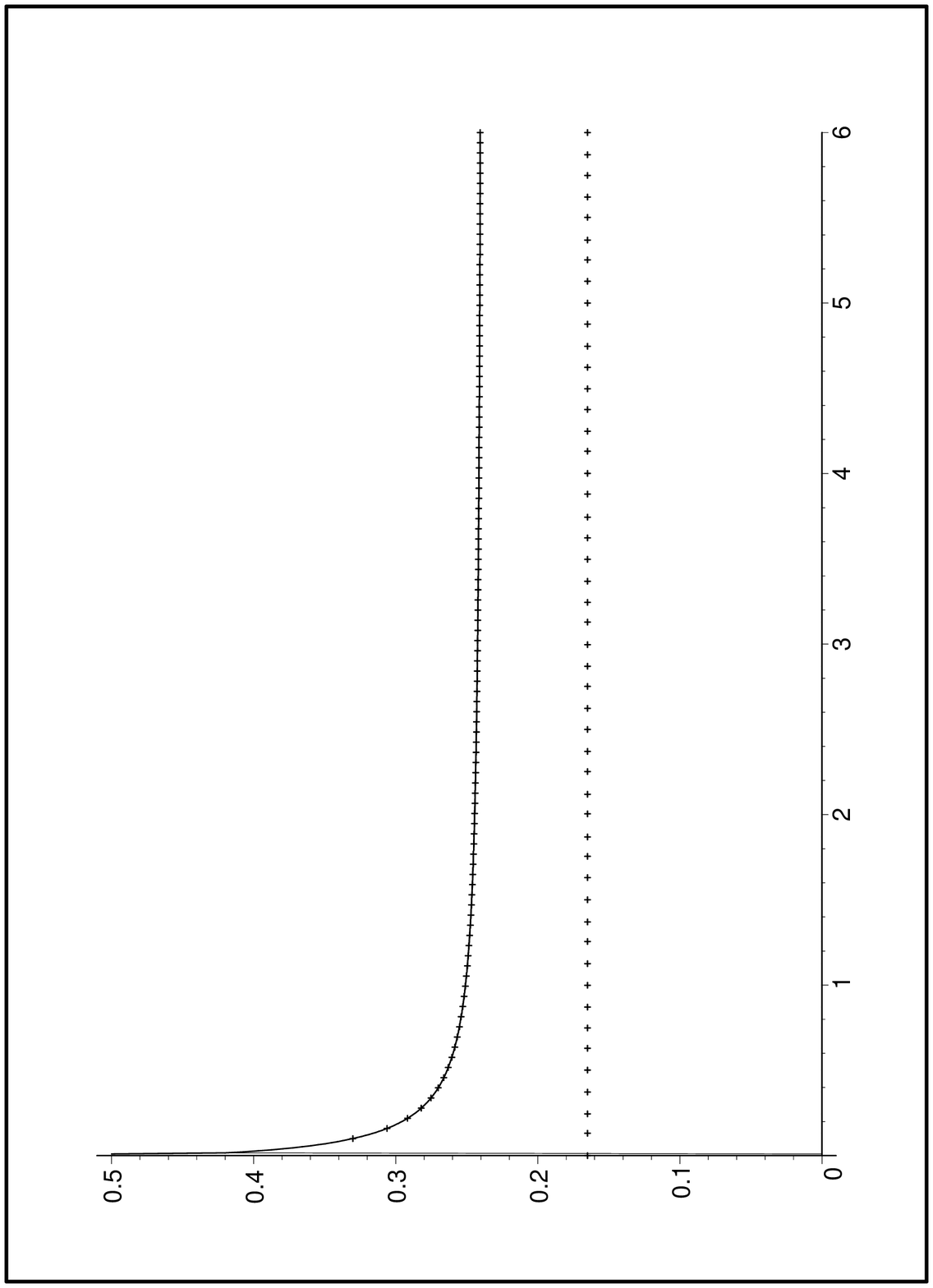}{\label{Ris6}An influence of the
neutrons' initial concentration on nucleosynthesis in the
none-equilibrium Universe at parameters' value
$\mathcal{P}_0=10^{18}$ and $\sigma_0=0,1$: along the abscissa
axis time in seconds is placed, along the ordinate axis - the
concentration of neutrons, $N_n(t)$ is placed. The thin line is
$N_n(0)=0$, heavy line is $N_n(0)=0,5$, dotted line is $N_n(0)=1$;
the value $N_n=0,165$ is noted via the dotted straight line. How
it is seen from the figure, all three curves practically coincide,
differing only for small values of time, $t<0,2$sec.}
The results of the numerical integration, shown on Fig.
\ref{Ris6}, confirm the fact, that the final concentration of
neutrons practically does not depend from their initial
concentrations. The number of neutrons after the hardening at
$\mathcal{P}_0\lesssim 10^{17}$ practically coincide with the
similar number in the hot model (SCS). The final outcome of
neutrons after the hardening at $\mathcal{P}_0>10^17$ increases
with the growth of $\mathcal{P}_0$, however at the constant value
of $\mathcal{P}_0$ slowly decreases with the growth of the
equilibrium parameter $\sigma_0$.

Adduced results of numerical calculations of the neutrons'
hardening in the non-equilibrium Universe, confirm, in general,
estimations, obtained earlier in papers \cite{Yuneq} and
\cite{Yudiffuz}, merely specifying their in details. Let us note,
that a maximal weight concentration of the cosmological He$_4$,
$N_{He_4}$, is equal to (see \cite{Zeld}):
\begin{equation}\label{116}
\max(N_{He_4})=2N_n.
\end{equation}
\REFigure{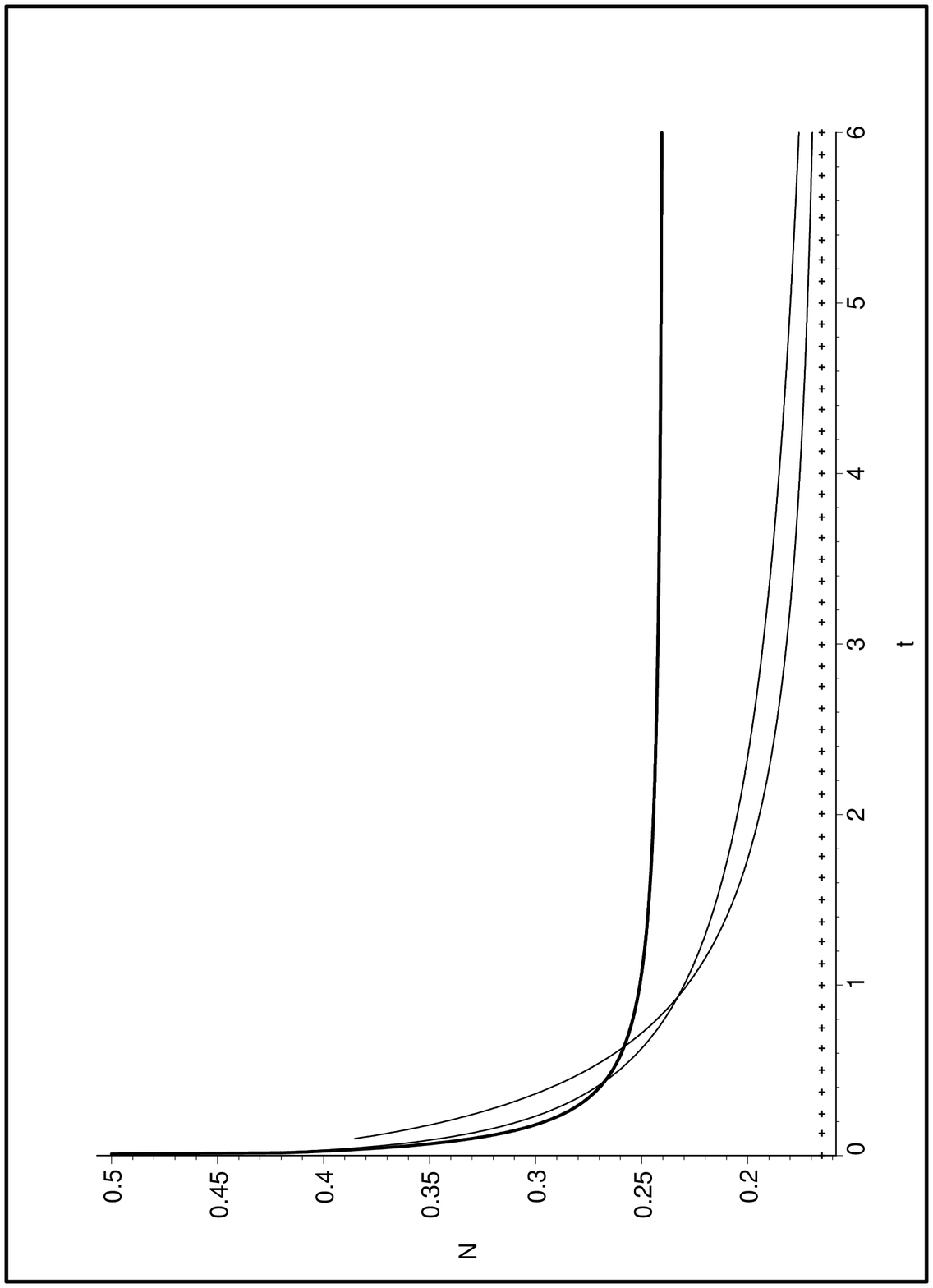}{\label{Ris7}An influence of the parameter
$\mathcal{P}_0$ on the nucleosynthesis in the non-equilibrium
Universe subject to the nucleosynthesis in the non-equilibrium
Universe at $N_n(0)=0,5$ and $\sigma_0=0,1$: on the abscissa axis
time in seconds is placed, on the ordinate axis the concentration
of neutrons, $N_n(t)$ is placed. The lower line is a classic
result (see \cite{Zeld}), the middle line is
$\mathcal{P}_0=10^{17}$, the heavy line is
$\mathcal{P}_0=10^{18}$; by means of the dotted line the value
$N_n=0,165$ is noted.}

\REFigure{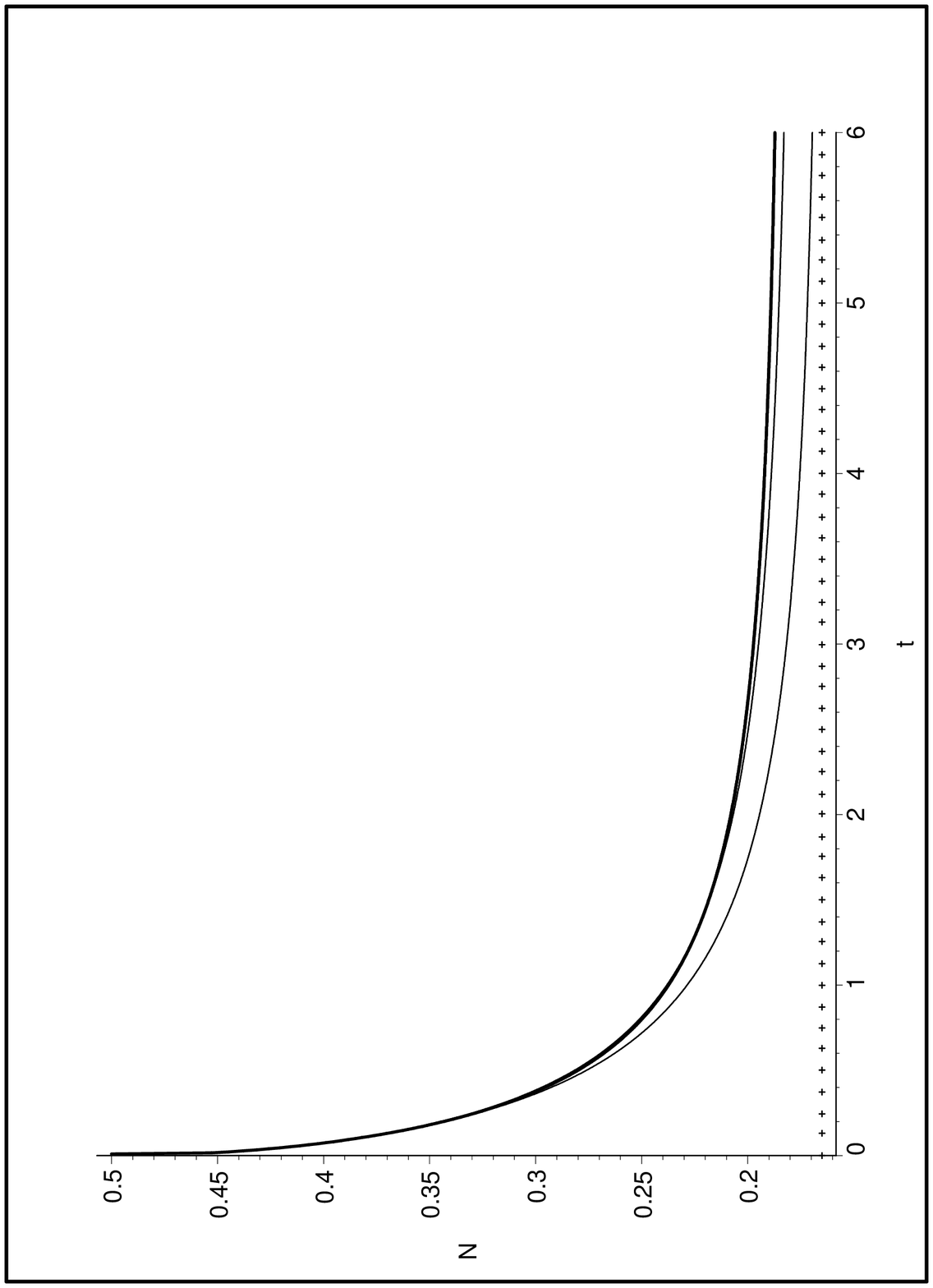}{\label{Ris8}An influence of the parameter
$\mathcal{P}_0$ on the nucleosynthesis in the non-equilibrium
Universe at $N_n(0)=0,5$ and $\sigma_0=0,9$: on the abscissa axis
time in seconds is placed, on the ordinate axis the concentration
of neutrons, $N_n(t)$ is placed. The lower line is a classic
result (see \cite{Zeld}), the middle line is
$\mathcal{P}_0=10^{17}$, the heavy line is
$\mathcal{P}_0=10^{18}$; by means of the dotted line the value
$N_n=0,165$ is noted.}

\REFigure{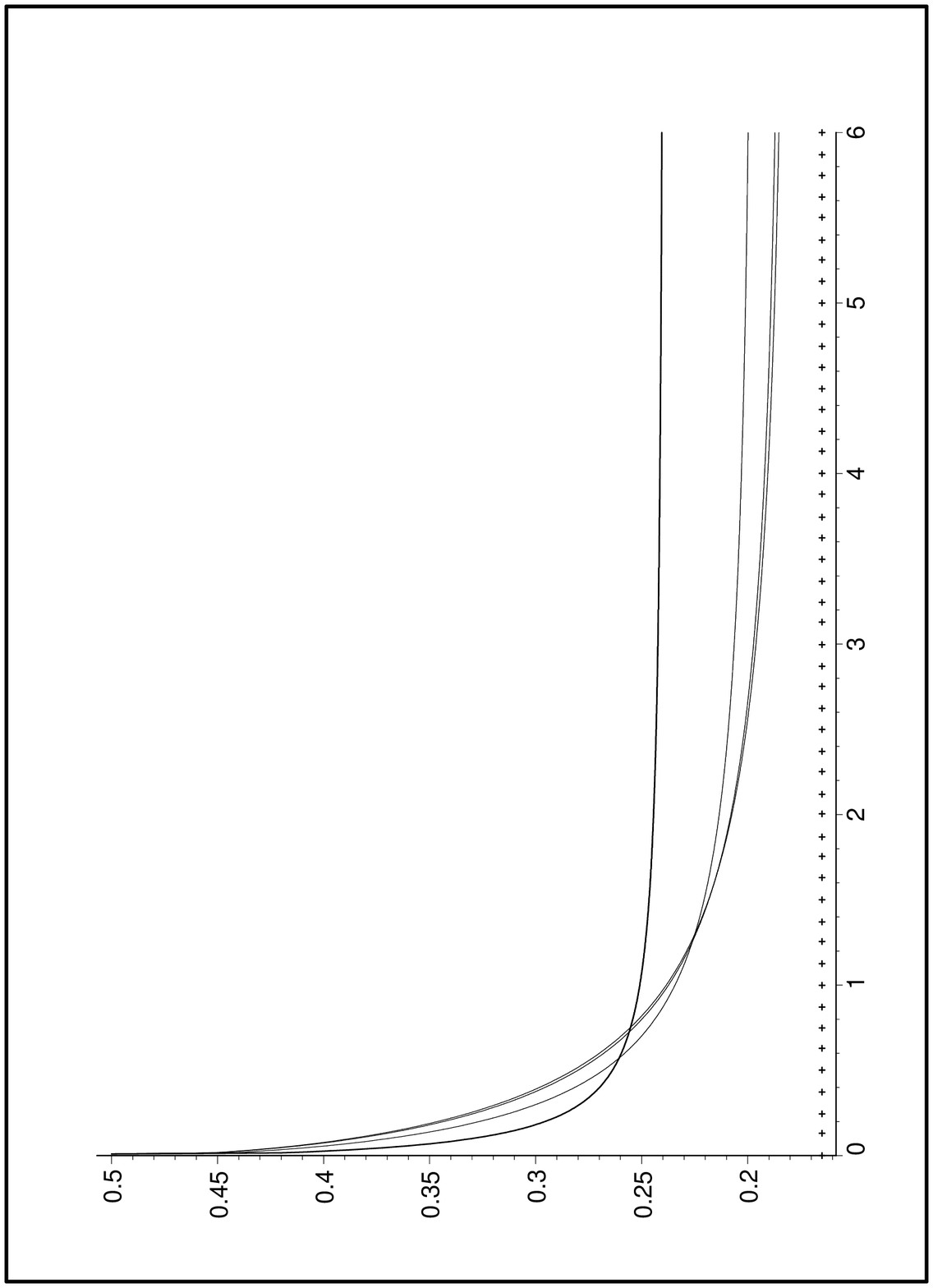}{\label{Ris9}An influence of the parameter
$\sigma_0$ on the nucleosynthesis in the non-equilibrium Universe
at $N_n(0)=0,5$ and $\mathcal{P}_0=10^{18}$: on the abscissa axis
time in seconds is placed, on the ordinate axis the concentration
of neutrons, $N_n(t)$ is placed. Top-down: $\sigma_0=0,1$,
$\sigma_0=0,5$, $\sigma_0=0,9$, $\sigma_0=0,99$; by means of the
dotted line the value $N_n=0,165$ is noted. Last two lines are
practically indistinguishable.}

\subsection{An observation restrictions on parameters of the non-equilibrium distribution}
Owing to the concentration growth of the cosmological $He_4$ an
observation data about the content of $He_4$ in the Universe on
the level not higher than 35\% at increase of the parameter
$\mathcal{P}_0$ imposes restrictions on upper values of the
non-equilibrium model's parameter $\mathcal{P}_0$ at a specified
value of its another parameter $\sigma_0$. This fact was noted
earlier in papers \cite{Yuneq}, \cite{Yudiffuz}, where on the
basis of quantitative assessments the range of allowed values of
non-equilibrium model's parameters was specified. However, these
values, correct at not large values of the parameter $\sigma_0$,
to be improved in the range $\sigma_0\sim 1$. Such improved
picture is represented on Fig. \ref{Ris10}.

\REFigure{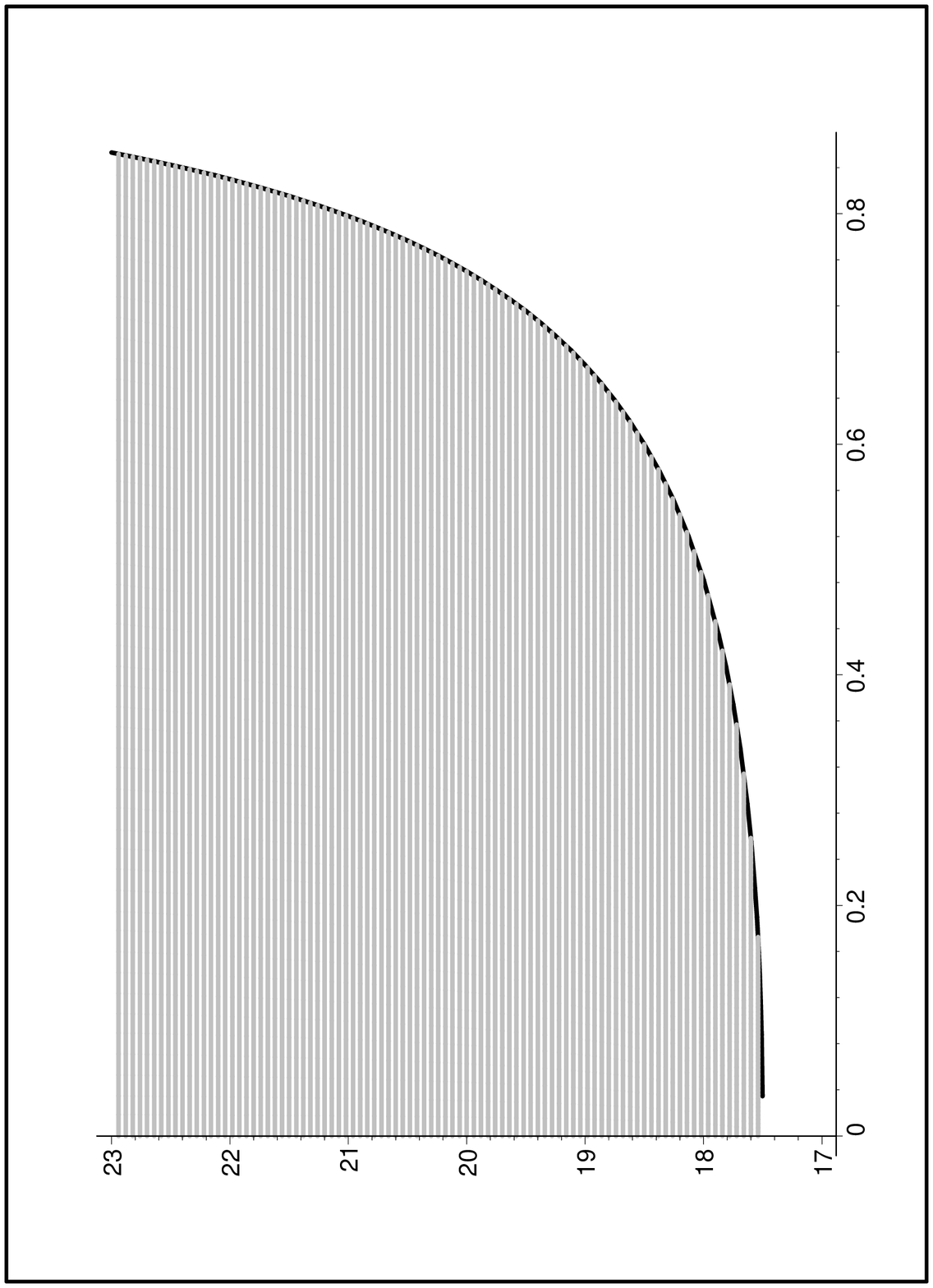}{\label{Ris10}An allowed by the concentration of
$He_4$ range of weakly-equilibrium model's parameters
$\lg\mathcal{P}_0$ (an ordinate axis) è $\sigma_0$ (an abscissa
axis) - an excluded range of parameters' values is hatched.}

\section*{Conclusion}
Summing up the paper's results, we note that modern data of an
observation cosmology do not contradict the supposition about a
weak initial violation of the Universe's thermal equilibrium in
terms of the condition (\ref{15}), i.e., to the initial smallness
of superthermal particles' number in comparison with the number of
thermal ones (see the condition (\ref{15}). An evolution of the
non-equilibrium Universe in the case of the initial distribution's
weak deviation from the thermal equilibrium, is managed generally
by two dimensionless parameters: a relative contribution to the
equilibrium component's energy density $0\leq\sigma_0\leq 1$ and a
ratio of the superthermal particles' average energy to the
temperature of the equilibrium Universe
$1<\overline{\mathcal{P}}_0<+\infty$ in the same point of time.
One could say also in such a way: the parameter
$\overline{\mathcal{P}}_0$ is equal to the non-equilibrium
particles' average energy in Planck scales of energy on Planck
point of time. An observation data at that, generally the
prevalence of the cosmological $He_4$, imposes certain
restrictions on the admitted range of weakly-equilibrium model's
parameters, but does not contradict values of the parameter
$\overline{\mathcal{P}}_0 \lesssim 3\cdot 10^{17}$, i.e. excesses
of an average energy of superthermal particles' component in 17(!)
orders of equilibrium components' temperature. Corresponding to
this parameter energy in the modern Universe at a temperature of
the relic radiation $3^oK$ is amount to $0,8*10^5$GeV(!). Our next
paper will be devoted to the investigation of the Universe model
with the initial strongly violated thermal equilibrium.

In conclusion authors express their thanks to professors V.N.
Melnikov, Yu.S. Vladimirov and D.V. Galtsov for a productive
discussion of paper's matters.

\end{document}